\documentclass[preprint]{imsart}

\usepackage{amsthm,amsmath}
\RequirePackage{natbib}
\RequirePackage[colorlinks,citecolor=blue,urlcolor=blue]{hyperref}

\arxiv{}

\startlocaldefs

\usepackage{amssymb,amsfonts,bbm,graphicx}
\usepackage[left=3.75cm,top=3.75cm,right=3.75cm,bottom=3.75cm]{geometry}

\usepackage{enumitem}
\newtheorem{lemma}{Lemma}
\newtheorem{corollary}{Corollary}

\theoremstyle{definition}
\newtheorem{definition}{Definition}

\theoremstyle{remark}

\makeatletter
\newtheorem*{rep@theorem}{\rep@title}
\newcommand{\newreptheorem}[2]{%
\newenvironment{rep#1}[1]{%
 \def\rep@title{#2 \ref{##1}}%
 \begin{rep@theorem}}%
 {\end{rep@theorem}}}
\makeatother

\newreptheorem{theorem}{Theorem}
\newreptheorem{lemma}{Lemma}
\newreptheorem{corollary}{Corollary}

\endlocaldefs

\begin{document}

\begin{frontmatter}

  \title{Statistical Properties of Convex Clustering} 
\runtitle{Convex Clustering}

\author{\fnms{Kean Ming}
  \snm{Tan}\corref{}\ead[label=e1]{keanming@uw.edu}} 
\address{Department of Biostatistics\\ University of Washington\\
Seattle, WA 98195, U.S.A.\\ \printead{e1}}
\and 
\author{\fnms{Daniela} \snm{Witten}\ead[label=e2]{dwitten@uw.edu}}
\address{Department of Statistics and Biostatistics \\ University of Washington\\
  Seattle, WA 98195, U.S.A.\\ \printead{e2}}

\runauthor{K.M.~Tan and D.~Witten}

\begin{abstract}
  In this manuscript, we study the statistical properties of \emph{convex clustering}. 
We establish that convex clustering is closely related to single linkage hierarchical clustering and $k$-means clustering. 
    In addition, we derive  the range of the tuning parameter for convex clustering that yields a non-trivial solution. 
     We also provide an unbiased estimator of the degrees of freedom, 
     and provide a finite sample bound for the prediction error  for convex clustering. 
      We compare convex clustering to some traditional clustering methods in simulation studies.
\end{abstract}


\end{frontmatter}

\section{Introduction} 
\label{Sec:intro}
Let $\mathbf{X}\in \mathbb{R}^{n\times p}$ be a data matrix with $n$ observations and $p$ features. We assume for convenience that the rows of $\mathbf{X}$ are unique.  The goal of clustering is to partition the $n$ observations into $K$ clusters, $D_1,\ldots,D_K$, based on some similarity measure.  Traditional clustering methods such as hierarchical clustering, $k$-means clustering, and spectral clustering take a greedy approach (see, e.g., \citealp{ElemStatLearn}).

In recent years, several authors have proposed formulations for \emph{convex clustering} \citep{pelckmans2005convex,hocking2011clusterpath,lindsten2011clustering,chi2013splitting}.  
\citet{chi2013splitting} proposed efficient algorithms for convex clustering.    In addition, \citet{radchenko2014consistent} studied the theoretical properties of a closely related problem to convex clustering, and \citet{NIPS2014_5307} studied the condition needed for convex clustering to recover the correct clusters.

 Convex clustering of the rows, $\mathbf{X}_{1.},\ldots,\mathbf{X}_{n.}$, of a data matrix $\mathbf{X}$ involves  solving the convex optimization problem
\begin{equation}
\label{Eq:convex clustering}
\underset{\mathbf{U}\in \mathbb{R}^{n\times p}}{\mathrm{minimize}} \; \frac{1}{2}\sum_{i=1}^n \|\mathbf{X}_{i.}- \mathbf{U}_{i.}\|^2_2 + \lambda \mathrm{Q}_q (\mathbf{U}),
\end{equation}  
where $\mathrm{Q}_q(\mathbf{U}) = \sum_{i<i'}\|\mathbf{U}_{i.}-\mathbf{U}_{i'.}\|_q$ for $q\in\{1,2, \infty\}$. The penalty  $\mathrm{Q}_{q}(\mathbf{U})$ generalizes the fused lasso penalty proposed in \citet{tibshirani2005sparsity}, and encourages the rows of $\hat{\mathbf{U}}$, the solution to~(\ref{Eq:convex clustering}), to take on a small number of unique values.  On the basis of $\hat{\mathbf{U}}$, we define the estimated clusters as follows.

\begin{definition}
\label{definition:cluster}
The $i$th and $i'$th observations are estimated by convex clustering to belong to the same cluster if and only if $\hat{\mathbf{U}}_{i.} = \hat{\mathbf{U}}_{i'.}$.  
\end{definition}

The tuning parameter $\lambda$ controls the number of unique rows of $\hat{\mathbf{U}}$, i.e., the number of estimated clusters.  When $\lambda = 0$, $\hat{\mathbf{U}}=\mathbf{X}$, and so each observation belongs to its own cluster.    As $\lambda$ increases, the number of unique rows of $\hat{\mathbf{U}}$ will decrease.  For sufficiently large $\lambda$, all rows of $\hat{\mathbf{U}}$ will be identical, and so all observations will be estimated to  belong to a single cluster.  
Note that (\ref{Eq:convex clustering}) is strictly convex, and therefore the solution $\hat{\mathbf{U}}$ is unique. 

To simplify our analysis of convex clustering, we rewrite (\ref{Eq:convex clustering}).
Let $\mathbf{x} = \mathrm{vec}(\mathbf{X}) \in  \mathbb{R}^{np}$ and let $\mathbf{u} = \mathrm{vec}(\mathbf{U}) \in \mathbb{R}^{np}$, where the $\mathrm{vec}(\cdot)$ operator is such that $x_{(i-1)p+j} = X_{ij}$ and $u_{(i-1)p+j} = U_{ij}$.  Construct $\mathbf{D}\in \mathbb{R}^{\left[p \cdot {n\choose 2}\right] \times np}$, and define the index set $\mathcal{C}(i,i')$ such that  the $p\times np$ submatrix $\mathbf{D}_{\mathcal{C}(i,i')}$ satisfies $\mathbf{D}_{\mathcal{C}(i,i')}\mathbf{u} = \mathbf{U}_{i.}-\mathbf{U}_{i'.}$.   Furthermore, for a vector $\mathbf{b}\in \mathbb{R}^{p \cdot {n\choose 2 }}$, we define 
\begin{equation}
\label{q norm}
\mathrm{P}_q (\mathbf{b}) = \sum_{i<i'} \|\mathbf{b}_{\mathcal{C}(i,i')}\|_q .
\end{equation}  
Thus, we have $\mathrm{P}_{q} (\mathbf{D}\mathbf{u}) = \sum_{ i<i'}\|\mathbf{D}_{\mathcal{C}(i,i')}
\mathbf{u}\|_q = \sum_{i <i'} \|\mathbf{U}_{i.}-\mathbf{U}_{i'.}\|_q = \mathrm{Q}_{q}(\mathbf{U})$.
Problem~(\ref{Eq:convex clustering}) can be rewritten as 
\begin{equation}
\label{Eq:convex clustering2}
\underset{\mathbf{u}\in \mathbb{R}^{np}}{\mathrm{minimize}} \; \frac{1}{2}\|\mathbf{x-u}\|_2^2 + \lambda \mathrm{P}_{q} (\mathbf{D}\mathbf{u}).
\end{equation}
When $q=1$, (\ref{Eq:convex clustering2}) is an instance of the  generalized lasso problem studied in \citet{tibshirani2011solution}.  Let $\hat{\mathbf{u}}$ be the solution to (\ref{Eq:convex clustering2}).  By Definition~\ref{definition:cluster}, the $i$th and $i'$th observations belong to the same cluster if and only if $\mathbf{D}_{\mathcal{C}(i,i')} \hat{\mathbf{u}}=0$.  In what follows, we work with (\ref{Eq:convex clustering2}) instead of (\ref{Eq:convex clustering}) for convenience.

 Let $\mathbf{D}^{\dagger}\in \mathbb{R}^{np \times \left[p\cdot {n\choose 2}\right]}$ be the Moore-Penrose pseudo-inverse of $\mathbf{D}$. We  state some properties of $\mathbf{D}$ and $\mathbf{D}^{\dagger}$ that will prove useful in later sections.  
\begin{lemma}
\label{lemma:properties of D}
The matrices $\mathbf{D}$ and $\mathbf{D}^{\dagger}$ have the following properties. 
\begin{enumerate}[label=(\roman*)]
\item \label{(i)} $\mathrm{rank}(\mathbf{D})= p(n-1)$.
\item \label{(ii)} $\mathbf{D}^{\dagger} = \frac{1}{n}\mathbf{D}^T$. 
\item \label{(iii)} $(\mathbf{D}^T\mathbf{D})^{\dagger} \mathbf{D}^T = \mathbf{D}^{\dagger}$ and $(\mathbf{DD}^T)^{\dagger} \mathbf{D} = (\mathbf{D}^T)^{\dagger}$.
\item \label{(iv)} $\mathbf{D}(\mathbf{D}^T\mathbf{D})^{\dagger}\mathbf{D}^T = \frac{1}{n}\mathbf{D}\mathbf{D}^T$ is a projection matrix onto the column space of $\mathbf{D}$.
\item \label{(v)} Define $\Lambda_{\min} (\mathbf{D})$ and $\Lambda_{\max} (\mathbf{D})$ as the minimum non-zero singular value and maximum singular value of the matrix $\mathbf{D}$, respectively.  Then, $ \Lambda_{\min} (\mathbf{D}) =\Lambda_{\max} (\mathbf{D}) = \sqrt{n}$.

\end{enumerate}
\end{lemma}

In this manuscript, we study the statistical properties of convex clustering.  In Section~\ref{Sec:motivation}, we study the dual problem of (\ref{Eq:convex clustering2}) and use it to establish that convex clustering is closely related to single linkage hierarchical clustering. In addition, we establish a connection between $k$-means clustering and convex clustering.  In Section~\ref{Sec:properties}, we present some properties of convex clustering.  More specifically, we characterize the range of the tuning parameter $\lambda$ in (\ref{Eq:convex clustering2}) such that convex clustering yields a non-trivial solution.  We also provide a finite sample bound for the prediction error, and an unbiased estimator of the degrees of freedom for convex clustering.  In Section~\ref{Sec:sim study}, we conduct numerical studies to evaluate the empirical performance of convex clustering relative to some existing proposals.  We close with a discussion in Section~\ref{Sec:discussion}.

\section{Convex Clustering, Single Linkage Hierarchical Clustering, and $k$-means Clustering}
\label{Sec:motivation}
In Section~\ref{subsec:dual}, we study the dual problem of convex clustering~(\ref{Eq:convex clustering2}).  Through its dual problem, we establish a connection between convex clustering and single linkage hierarchical  clustering in Section~\ref{subsec:connection slc}. We then show that convex clustering is closely related to $k$-means clustering in Section~\ref{subsec:kmeans}.

\subsection{Dual Problem of Convex Clustering}
\label{subsec:dual}
We analyze convex clustering~(\ref{Eq:convex clustering2}) by studying its dual problem.  Let $s,q\in \{1,2,\infty\}$ satisfy $\frac{1}{s}+\frac{1}{q}=1$.   For a vector $\mathbf{b} \in \mathbb{R}^{p\cdot {n\choose 2}}$, let $\mathrm{P}^*_q (\mathbf{b})$ denote the dual norm of $\mathrm{P}_q(\mathbf{b})$, which takes the form
\begin{equation}
\label{dual norm}
\mathrm{P}_q^* (\mathbf{b}) = \underset{i<i'}{\max}\; \| \mathbf{b}_{\mathcal{C}(i,i')}\|_s.
\end{equation}
  We refer the reader to Chapter 6 in \citet{boyd2004convex} for an overview of the concept of duality.

\begin{lemma}
\label{lemma:dual problem}
The dual problem of convex clustering~(\ref{Eq:convex clustering2}) is 
\begin{equation}
\label{Eq:convex dual}
\underset{\boldsymbol{\nu}\in\mathbb{R}^{\left[p \cdot {n\choose 2}\right]}}{\mathrm{minimize}} \; \frac{1}{2}\|\mathbf{x}-\mathbf{D}^T\boldsymbol{\nu}\|_2^2 \quad \quad \mathrm{subject \; to}\; \mathrm{P}^*_q(\boldsymbol{\nu}) \le \lambda,
\end{equation}
where $\boldsymbol{\nu} \in \mathbb{R}^{\left[p\cdot {n\choose 2}\right]}$ is the dual variable. Furthermore, let $\hat{\mathbf{u}}$ and $\hat{\boldsymbol{\nu}}$ be the solutions to (\ref{Eq:convex clustering2}) and (\ref{Eq:convex dual}), respectively.  Then, 
\begin{equation}
\label{Eq:convex clustering2 solution}
\mathbf{D}\hat{\mathbf{u}} = \mathbf{D}\mathbf{x}-\mathbf{D} \mathbf{D}^T\hat{\boldsymbol{\nu} }.
\end{equation}
\end{lemma}
 While (\ref{Eq:convex clustering2}) is strictly convex, its dual problem~(\ref{Eq:convex dual}) is not strictly convex, since $\mathbf{D}$ is not of full rank by Lemma~\ref{lemma:properties of D}\ref{(i)}.  
Therefore, the solution $\hat{\boldsymbol{\nu}}$ to (\ref{Eq:convex dual}) is not unique.
Lemma~\ref{lemma:properties of D}\ref{(iv)} indicates that $\frac{1}{n}\mathbf{DD}^T$ is a projection matrix onto the column space of $\mathbf{D}$.  Thus, the solution $\mathbf{D}\hat{\mathbf{u}}$ in (\ref{Eq:convex clustering2 solution}) can be interpreted as the difference between $\mathbf{Dx}$, the pairwise difference between rows of $\mathbf{X}$, and the projection of a dual variable onto the column space of $\mathbf{D}$.

We now consider a modification to the convex clustering problem (\ref{Eq:convex clustering2}). Recall from Definition~\ref{definition:cluster}  that the $i$th and $i$'th observations are in the same estimated cluster if  $\mathbf{D}_{\mathcal{C}(i,i')}\hat{\mathbf{u}}=\mathbf{0}$.  This motivates us to estimate $\boldsymbol{\gamma}=\mathbf{Du}$ directly by solving
\begin{equation}
\label{Eq:convex modification}
\underset{\boldsymbol{\gamma}\in \mathbb{R}^{\left[p \cdot {n\choose 2}\right]}}{\mathrm{minimize}} \; \frac{1}{2}\|\mathbf{Dx}-\boldsymbol{\gamma}\|_2^2 + \lambda \mathrm{P}_{q} (\boldsymbol{\gamma}).
\end{equation}
We establish  a connection between  (\ref{Eq:convex clustering2}) and  (\ref{Eq:convex modification}) by studying the dual problem of~(\ref{Eq:convex modification}). 
\begin{lemma}
\label{lemma:dual problem2}
The dual problem of (\ref{Eq:convex modification}) is 
\begin{equation}
\label{Eq:convex modification dual}
\underset{\boldsymbol{\nu}'\in\mathbb{R}^{\left[p \cdot {n\choose 2}\right]}}{\mathrm{minimize}} \; \frac{1}{2}\|\mathbf{Dx}-\boldsymbol{\nu}'\|_2^2 \quad \quad \mathrm{subject \; to \;} \mathrm{P}^*_q(\boldsymbol{\nu}') \le \lambda,
\end{equation}
where $\boldsymbol{\nu}'\in\mathbb{R}^{\left[p \cdot {n\choose 2}\right]}$ is the dual variable.  Furthermore, let $\hat{\boldsymbol{\gamma}}$ and $\hat{\boldsymbol{\nu}}'$ be the solutions to (\ref{Eq:convex modification}) and (\ref{Eq:convex modification dual}), respectively.  Then, 
\begin{equation}
\label{Eq:sol modification}
\hat{\boldsymbol{\gamma}} = \mathbf{Dx} - \hat{\boldsymbol{\nu}}'.
\end{equation}
\end{lemma}
Comparing~(\ref{Eq:convex clustering2 solution}) and~(\ref{Eq:sol modification}), we see that the solutions to convex clustering~(\ref{Eq:convex clustering2}) and the modified problem~(\ref{Eq:convex modification}) are closely related.
  In particular, both $\mathbf{D}\hat{\mathbf{u}}$ in~(\ref{Eq:convex clustering2 solution}) and $\hat{\boldsymbol{\gamma}}$ in~(\ref{Eq:sol modification}) 
    involve taking the difference between $\mathbf{Dx}$ and some function of a dual variable that has $\mathrm{P}^*_q(\cdot)$ norm less than or equal to $\lambda$.  
The main difference is that in (\ref{Eq:convex clustering2 solution}), the dual variable is projected into the column space of $\mathbf{D}$.

Problem (\ref{Eq:convex modification}) is quite simple, and in fact it amounts to a thresholding operation on $\mathbf{Dx}$ when $q=1$ or $q=2$, i.e., the solution $\hat{\boldsymbol{\gamma}}$ is obtained by performing soft thresholding on $\mathbf{D}\mathbf{x}$, or group soft thresholding on $\mathbf{D}_{\mathcal{C}(i,i')}\mathbf{x}$ for all $i<i'$, respectively \citep{bach2011convex}.  When $q=\infty$, an efficient algorithm was proposed by \citet{duchisinger2009}.    

\subsection{Convex Clustering and Single Linkage Hierarchical Clustering}
\label{subsec:connection slc}

In this section, we establish a connection between convex clustering and single linkage hierarchical clustering.  
Let $\hat{\boldsymbol{\gamma}}^q$ be the solution to (\ref{Eq:convex modification}) with $\mathrm{P}_q(\cdot)$ norm and let $s,q\in \{1,2,\infty\}$ satisfy $\frac{1}{s}+\frac{1}{q} = 1$.   Since~(\ref{Eq:convex modification}) is separable in $\boldsymbol{\gamma}_{\mathcal{C}(i,i')}$ for all $i<i'$, by Lemma 2.1 in \citet{haris2014convex}, it can be verified that
 \begin{equation}
 \label{Eq:gammahat new}
 \hat{\boldsymbol{\gamma}}_{\mathcal{C}(i,i')}^q=\mathbf{0} \quad \mathrm{ if\; and\; only\; if} \quad  
 \|\mathbf{X}_{i.}-\mathbf{X}_{i'.}\|_s \le \lambda.
 \end{equation}
It might be tempting to conclude that a pair of observations $(i,i')$ belong to the same cluster if $\hat{\boldsymbol{\gamma}}_{\mathcal{C}(i,i')}^q=\mathbf{0}$.  However, by inspection of (\ref{Eq:gammahat new}), it could happen that $\hat{\boldsymbol{\gamma}}_{\mathcal{C}(i,i')}^q = \mathbf{0}$ and $\hat{\boldsymbol{\gamma}}_{\mathcal{C}(i',i'')}^q = \mathbf{0}$, but $\hat{\boldsymbol{\gamma}}_{\mathcal{C}(i,i'')}^q \ne \mathbf{0}$.

To overcome this problem, we define the $n\times n$ adjacency matrix $\mathbf{A}^q(\lambda)$ as
\begin{equation}
\label{Eq:adjacency}
\left[\mathbf{A}^{q} (\lambda)  \right]_{ii'}= \begin{cases}  1& \text{if } i=i', \\ 1 & \text{if }\hat{\boldsymbol{\gamma}}_{\mathcal{C}(i,i')}^q=\mathbf{0},\\
 0 & \text{if }\hat{\boldsymbol{\gamma}}_{\mathcal{C}(i,i')}^q\ne\mathbf{0}.
\end{cases}
\end{equation}
Subject to a rearrangement of the rows and columns, $\mathbf{A}^q(\lambda)$ is a block-diagonal matrix with some number of blocks, denoted as $R$.  On the basis of $\mathbf{A}^q(\lambda)$, we define $R$ estimated clusters: the indices of the observations in the $r$th cluster are the same as the indices of the observations in the $r$th block of $\mathbf{A}^q(\lambda)$.

We now present a lemma on the equivalence between single linkage hierarchical clustering and the clusters identified by (\ref{Eq:convex modification}) using (\ref{Eq:adjacency}). The lemma follows directly from the definition of single linkage clustering (see, for instance, Chapter 3.2 of \citealp{JD88}).
\begin{lemma} 
\label{lemma:connection to SLC}
 Let $\hat{E}_1,\ldots,\hat{E}_R$ index the blocks within the adjacency matrix $\mathbf{A}_q(\lambda)$. Let $s$ satisfy $\frac{1}{s}+\frac{1}{q}=1$.     Let $\hat{D}_1,\ldots,\hat{D}_K$ denote the clusters that result from performing  single linkage hierarchical clustering on the dissimilarity matrix defined by the pairwise distance between the observations $\|\mathbf{X}_{i.}-\mathbf{X}_{i'}\|_{s}$, and cutting the dendrogram at the height of $\lambda>0$.    Then $K=R$, and there exists a permutation $\pi: \{1,\ldots,K\} \rightarrow \{1,\ldots,K\}$ such that $D_k=E_{\pi(k)}$ for $k=1,\ldots,K$.
\end{lemma}
\noindent In other words, Lemma~\ref{lemma:connection to SLC} implies that  single linkage hierarchical clustering and (\ref{Eq:convex modification}) yield the same estimated clusters.  Recalling the connection between (\ref{Eq:convex clustering2}) and (\ref{Eq:convex modification}) established in Section~\ref{subsec:dual}, this implies a close connection between convex clustering and single linkage hierarchical clustering.

\subsection{Convex Clustering and $k$-Means Clustering}
\label{subsec:kmeans}
We now establish a connection between convex clustering and $k$-means clustering.
$k$-means clustering seeks to partition the $n$ observations into $K$ clusters by minimizing the within cluster sum of squares.  That is, the clusters are given by the partition $\hat{D}_1,\ldots,\hat{D}_{K}$ of $\{1,\ldots,n\}$ that solves the optimization problem
\begin{equation}
\label{Eq:kmeans}
\underset{\boldsymbol{\mu}_1,\ldots,\boldsymbol{\mu}_K \in \mathbb{R}^p,D_1,\ldots,D_K}{\mathrm{minimize}}\;  \sum_{k=1}^K \sum_{i\in D_k} \|\mathbf{X}_{i.}- \boldsymbol{\mu}_k  \|_2^2.
\end{equation}

We  consider convex clustering~(\ref{Eq:convex clustering}) with $q=0$,
\begin{equation}
\label{Eq:cc-kmeans}
\underset{\mathbf{U}\in \mathbb{R}^{n\times p}}{\mathrm{minimize}} \; \frac{1}{2}\sum_{i=1}^n \|\mathbf{X}_{i.}- \mathbf{U}_{i.}\|^2_2 + \lambda \sum_{i<i'} \mathbb{I} (\mathbf{U}_{i.}\ne \mathbf{U}_{i'.}),
\end{equation}
where $\mathbb{I} (\mathbf{U}_{i.} \ne\mathbf{U}_{i'.})$ is an indicator function that equals one if  $ \mathbf{U}_{i.}\ne \mathbf{U}_{i'.}$.   Note that~(\ref{Eq:cc-kmeans}) is no longer a convex optimization problem.

We now establish a connection between~(\ref{Eq:kmeans}) and~(\ref{Eq:cc-kmeans}).
For a given value of $\lambda$, 
(\ref{Eq:cc-kmeans}) is equivalent to  
\begin{equation}
\label{Eq:cc-kmeans3}
\underset{\mathbf{U}\in \mathbb{R}^{n\times p}, K, \boldsymbol{\mu}_{1},\ldots,\boldsymbol{\mu}_K\in \mathbb{R}^p,E_1,\ldots,E_K}{\mathrm{minimize}} \; \frac{1}{2}\sum_{k=1}^K \sum_{i\in E_k} \|\mathbf{X}_{i.}- \boldsymbol{\mu}_{k}\|^2_2  + \lambda \sum_{i<i'} \sum_{k=1}^K \mathbb{I} (i\in E_k, i' \notin E_{k}),
\end{equation}
subject to the constraint that $\{\boldsymbol{\mu}_1,\ldots,\boldsymbol{\mu}_K\}$ are the unique rows of $\mathbf{U}$ and $E_k = \{i: \mathbf{U}_{i.}=\boldsymbol{\mu}_k\}$.  Note that  $\mathbb{I} (i\in E_k, i' \notin E_{k})$ is an indicator function that equals to one if $i \in E_k$ and $i' \notin E_{k}$.
Thus, we see from (\ref{Eq:kmeans}) and (\ref{Eq:cc-kmeans3}) that  $k$-means clustering is equivalent to convex clustering with $q=0$, up to a penalty term $ \lambda \sum_{i<i'} \sum_{k=1}^K \mathbb{I} (i\in E_k, i' \notin E_{k})$.

To interpret the penalty term, we consider the case when there are two clusters $E_1$ and $E_2$.  The penalty term reduces to $\lambda |E_1| \cdot (n-|E_1|)$, where $|E_1|$ is the cardinality of the set $E_1$.  The term $\lambda |E_1| \cdot (n-|E_1|)$ is minimized when $|E_1|$ is either 1 or $n-1$, encouraging one cluster taking only one observation. 
Thus, compared to $k$-means clustering, convex clustering with $q=0$ has the undesirable behavior of producing clusters whose sizes are highly unbalanced.

\section{Properties of Convex Clustering}
\label{Sec:properties}
We now study the properties of convex clustering~(\ref{Eq:convex clustering2}) with $q\in \{1,2\}$.   In Section~\ref{Sec:complete sparsity}, we establish the range of the tuning parameter $\lambda$ in (\ref{Eq:convex clustering2}) such that convex clustering yields a non-trivial solution with more than one cluster.  We provide finite sample bounds for the prediction error  of convex clustering in Section~\ref{Sec:prediction}. Finally, we provide unbiased estimates of the degrees of freedom for convex clustering  in Section~\ref{Sec:degrees of freedom}. 

\subsection{Range of $\lambda$ that Yields Non-trivial Solution}
\label{Sec:complete sparsity}
In this section, we establish the range of the tuning parameter $\lambda$ such that convex clustering (\ref{Eq:convex clustering2}) yields a solution with more than one cluster.  
\begin{lemma}
\label{lemma:max lambda}
Let 
\begin{equation}
\label{Eq:lambda range}
\lambda_{\mathrm{upper}} := \begin{cases} \underset{\boldsymbol{\omega}}{\min} \;  \left\| \frac{1}{n}\mathbf{Dx} + \left(\mathbf{I} - \frac{1}{n}\mathbf{D}\mathbf{D}^T\right) {\boldsymbol{\omega}}\right\|_\infty & \mathrm{for \;} q=1,\\
\underset{\boldsymbol{\omega}}{\min} \;  \left\{ \underset{i<i'}{\max} \;   \left\{ \left\| \left(\frac{1}{n}\mathbf{Dx} + \left(\mathbf{I} - \frac{1}{n}\mathbf{D}\mathbf{D}^T\right) {\boldsymbol{\omega}}\right)_{\mathcal{C}(i,i')}\right\|_2\right\} \right\}  & \mathrm{for\; } q=2.\end{cases}
\end{equation}
Convex clustering (\ref{Eq:convex clustering2}) with $q=1$ or $q=2$ yields  a non-trivial solution of  more than one cluster if and only if  $\lambda < \lambda_{\mathrm{upper}}$.
\end{lemma}
\noindent By Lemma~\ref{lemma:max lambda}, we see that calculating $\lambda_{\mathrm{upper}}$ boils down to solving a convex optimization problem. This can be solved using a standard solver such as \verb=CVX= in \verb=MATLAB=. In the absence of such a solver, a loose upper bound on $\lambda_{\mathrm{upper}}$ is given by $\|\frac{1}{n}\mathbf{Dx}\|_{\infty}$ for $q=1$, or $\underset{i<i'}{\max} \; \|\frac{1}{n}\mathbf{D}_{\mathcal{C}(i,i')}\mathbf{x}\|_2 $ for $q=2$.

Therefore, to obtain the entire solution path of convex clustering, we need only consider values of $\lambda$ that satisfy $\lambda \le \lambda_{\mathrm{upper}}$.

\subsection{Bounds on Prediction Error}
\label{Sec:prediction}
In this section, we assume the model 
$\mathbf{x} = \mathbf{u}+\boldsymbol{\epsilon}$, where $\boldsymbol{\epsilon} \in \mathbb{R}^{np}$ is a vector of independent sub-Gaussian noise terms with mean zero and variance $\sigma^2$, and $\mathbf{u}$ is an arbitrary $np$-dimensional mean vector.  We refer the reader to pages 24-25 in \citet{boucheron2013concentration} for the properties of sub-Gaussian random variables.
We now provide finite sample bounds for the prediction error of convex clustering (\ref{Eq:convex clustering2}).    Let $\lambda$ be the tuning parameter in (\ref{Eq:convex clustering2}) and let $\lambda' = \frac{\lambda }{np}$.

\begin{lemma}
\label{lemma:prediction consistency}
Suppose that $\mathbf{x}=\mathbf{u}+\boldsymbol{\epsilon}$, where $\boldsymbol{\epsilon} \in \mathbb{R}^{np}$ and the elements of $\boldsymbol{\epsilon}$ are independent sub-Gaussian random variables with mean zero and variance $\sigma^2$.  Let $\hat{\mathbf{u}}$ be the estimate obtained from (\ref{Eq:convex clustering2}) with $q=1$.  If $\lambda' \ge 4\sigma \sqrt{\frac{\log \left( p\cdot {n\choose 2}   \right)}{ n^3p^2}}$,
 then
\begin{equation*}
\frac{1}{2np} \|\hat{\mathbf{u}} - \mathbf{u} \|^2_2 \le  \frac{3\lambda'}{2} \|\mathbf{Du}\|_1  + \sigma^2 \left[\frac{1}{n} +  \sqrt{\frac{\log (np)}{n^2p}}  \right]
\end{equation*}
holds with probability at least $1-\frac{2}{p\cdot {n\choose 2}}-  \exp \left\{  -\min \left( c_1 \log (np), c_2 \sqrt{p\log (np)} \right) \right\}$, where $c_1$ and $c_2$ are positive constants appearing in Lemma~\ref{lemma:hanson wright}.
\end{lemma}
\noindent We see from Lemma~\ref{lemma:prediction consistency} that the average prediction error is bounded by the oracle quantity $\|\mathbf{Du}\|_1$ and a second term that decays to zero as $n,p \rightarrow \infty$.   Convex clustering with $q=1$ is prediction consistent only if $\lambda' \|\mathbf{Du}\|_1 = {o} \left(1  \right)$.  We now provide a scenario for which $\lambda' \|\mathbf{Du}\|_1 = {o} \left(1 \right)$ holds.

Suppose that we are in the high-dimensional setting in which $p>n$ and  the true underlying clusters differ only with respect to a fixed number of features \citep{witten2010framework}.  Also, suppose that each element of $\mathbf{Du}$ --- that is, $U_{ij}-U_{i'j}$ --- is of order $O(1)$.  Therefore, $\|\mathbf{Du}\|_1=O(n^2)$, since by assumption only a fixed number of  features have different means across clusters.  Assume that  $\sqrt{\frac{n \log \left(p \cdot {n\choose 2}\right)}{p^2}} ={o}(1)$.  Under these assumptions, convex clustering with $q=1$ is prediction consistent. 

Next, we present a finite sample bound on the prediction error for convex clustering with $q=2$.
\begin{lemma}
\label{lemma:prediction consistency l2}
Suppose that $\mathbf{x}=\mathbf{u}+\boldsymbol{\epsilon}$, where $\boldsymbol{\epsilon} \in \mathbb{R}^{np}$ and the elements of $\boldsymbol{\epsilon}$ are independent sub-Gaussian random variables with mean zero and variance $\sigma^2$.  Let $\hat{\mathbf{u}}$ be the estimate obtained from (\ref{Eq:convex clustering2}) with $q=2$.  If $\lambda' \ge 4\sigma \sqrt{\frac{\log \left( p\cdot {n\choose 2}   \right)}{ n^3p}}$,
 then
\[
\frac{1}{2np} \|\hat{\mathbf{u}} - \mathbf{u} \|^2_2 \le  \frac{3\lambda'}{2} \sum_{i<i'}\|\mathbf{D}_{\mathcal{C}(i,i')}\mathbf{u}\|_2  + \sigma^2 \left[\frac{1}{n} +  \sqrt{\frac{\log (np)}{n^2p}}   \right]
\]
holds with probability at least $1-\frac{2}{ p \cdot {n\choose 2}}-  \exp \left\{  -\min \left( c_1 \log (np), c_2 \sqrt{p\log (np)} \right) \right\}$, where $c_1$ and $c_2$ are positive constants appearing in Lemma~\ref{lemma:hanson wright}.
\end{lemma}
\noindent Under the scenario described above, $\|\mathbf{D}_{\mathcal{C}(i,i')}\mathbf{u}\|_2 = O(1)$, and therefore $\sum_{i<i'}\|\mathbf{D}_{\mathcal{C}(i,i')}\mathbf{u}\|_2  = O(n^2)$.   Convex clustering with $q=2$ is prediction consistent if $\sqrt{\frac{n \log \left(p \cdot {n\choose 2}\right)}{p}} ={o}(1)$.

\subsection{Degrees of Freedom}
\label{Sec:degrees of freedom}
Convex clustering recasts the clustering problem as a penalized regression problem, for which the notion of degrees of freedom is established \citep{efron1986biased}.  Under this framework, we provide an unbiased estimator of the degrees of freedom for clustering.  Recall that $\hat{\mathbf{u}}$   is the solution to convex clustering (\ref{Eq:convex clustering2}). Suppose that $\mathrm{Var}(\mathbf{x}) = \sigma^2 \mathbf{I}$.  Then, the degrees of freedom for convex clustering is defined as $\frac{1}{\sigma^2} \sum_\mathrm{j=1}^{np}\mathrm{Cov}(\hat{u}_j,x_j)$ (see, e.g., \citealp{efron1986biased}). An unbiased estimator of the degrees of freedom for convex clustering with $q=1$ follows directly from Theorem 3 in \citet{tibshirani2012degrees}.  
\begin{lemma}
\label{dof:l1}
Assume that $\mathbf{x} \sim \mathrm{MVN}(\mathbf{u},\sigma^2\mathbf{I})$, and let $\hat{\mathbf{u}}$ be the solution to (\ref{Eq:convex clustering2}) with $q=1$.  Furthermore, let $\hat{\mathcal{B}}_1 = \{j : (\mathbf{D}\hat{\mathbf{u}})_j\ne 0 \}$.  We define the matrix $\mathbf{D}_{-\hat{\mathcal{B}}_1}$ by removing the rows of $\mathbf{D}$ that correspond to $\hat{\mathcal{B}}_1$.  Then
\begin{equation}
\label{Eq:dof1}
\begin{split}
\hat{\mathrm{df}}_{1} &=\mathrm{tr} \left( \mathbf{I} - \mathbf{D}_{-\hat{\mathcal{B}}_1}^T (\mathbf{D}_{-\hat{\mathcal{B}}_1}\mathbf{D}_{-\hat{\mathcal{B}}_1}^T)^{\dagger} \mathbf{D}_{-\hat{\mathcal{B}}_1}   \right)\\
\end{split}
\end{equation}
is an unbiased estimator of the degrees of freedom of convex clustering with $q=1$.  
\end{lemma}
\noindent The following corollary follows directly from Corollary 1 in \citet{tibshirani2011solution}.
\begin{corollary}
Assume that $\mathbf{x} \sim \mathrm{MVN}(\mathbf{u},\sigma^2\mathbf{I})$, and let $\hat{\mathbf{u}}$ be the solution to (\ref{Eq:convex clustering2}) with $q=1$. The fit $\hat{\mathbf{u}}$ has degrees of freedom 
\[
\mathrm{df}_1(\hat{\mathbf{u}}) = \mathrm{E}\left[ \mathrm{number\; of\; unique\; elements\; in \;} \hat{\mathbf{u}}
\right].
\]
\end{corollary}
 There is an interesting interpretation of the degrees of freedom estimator for convex clustering with $q=1$.  Suppose that there are $K$ estimated clusters, and all elements of the estimated means corresponding to the $K$ estimated clusters are unique.  Then the degrees of freedom is $Kp$, the product of the number of estimated clusters and the number of features.
 
Next, we provide an unbiased estimator of the degrees of freedom for convex clustering with $q=2$.  
\begin{lemma}
\label{lemma:dof:l2}
 Assume that $\mathbf{x} \sim \mathrm{MVN}(\mathbf{u},\sigma^2\mathbf{I})$,  and let $\hat{\mathbf{u}}$ be the solution to (\ref{Eq:convex clustering2}) with $q=2$.  Furthermore, let $\hat{\mathcal{B}}_2 = \{(i,i') : \|\mathbf{D}_{\mathcal{C}(i,i')}\hat{\mathbf{u}}\|_2\ne 0 \}$. We define the matrix $\mathbf{D}_{-\hat{\mathcal{B}}_2}$ by removing rows of $\mathbf{D}$ that correspond to $\hat{\mathcal{B}}_2$.   Let $\mathbf{P} = \left( \mathbf{I}- \mathbf{D}^T_{-\hat{\mathcal{B}}_2}  (\mathbf{D}_{-\hat{\mathcal{B}}_2}  \mathbf{D}^T_{-\hat{\mathcal{B}}_2})^{\dagger} \mathbf{D}_{-\hat{\mathcal{B}}_2}     \right)$ be the projection matrix onto the complement of the space spanned by the rows of $\mathbf{D}_{-\hat{\mathcal{B}}_2}$.  Then
\begin{equation}
\label{Eq:dof2}
\hat{\mathrm{df}}_{2} =\mathrm{tr} \left(  \left[\mathbf{I}+ \lambda \mathbf{P} \sum_{(i,i')\in \hat{\mathcal{B}}_2 } \left( \frac{\mathbf{D}^T_{\mathcal{C}(i,i')} \mathbf{D}_{\mathcal{C}(i,i')} }{\|\mathbf{D}_{\mathcal{C}(i,i')}   \hat{\mathbf{u}}\|_2} -  \frac{\mathbf{D}^T_{\mathcal{C}(i,i')} \mathbf{D}_{\mathcal{C}(i,i')}   \hat{\mathbf{u}}\hat{\mathbf{u}}^T \mathbf{D}^T_{\mathcal{C}(i,i')} \mathbf{D}_{\mathcal{C}(i,i')}  }{\|\mathbf{D}_{\mathcal{C}(i,i')}   \hat{\mathbf{u}}\|_2^3}\right)\right]^{-1} \mathbf{P}\right)
\end{equation}
is an unbiased estimator of the degrees of freedom of convex clustering with $q=2$.  
\end{lemma}
\noindent When $\lambda = 0$, $\|\mathbf{D}_{\mathcal{C}(i,i')}\hat{\mathbf{u}}\|_2 \ne 0$ for all $i<i'$. Therefore, $\mathbf{P}=\mathbf{I} \in \mathbb{R}^{np\times np}$ and the degrees of freedom estimate is equal to $\mathrm{tr}(\mathbf{I}) = np$.    When $\lambda$ is sufficiently large that $\hat{\mathcal{B}}_2$ is an empty set, one can verify that $\mathbf{P}= \mathbf{I}- \mathbf{D}^T(\mathbf{D}\mathbf{D}^T)^\dagger\mathbf{D}$ is a projection matrix of rank $p$, using the fact that $\mathrm{rank}(\mathbf{D}) = p(n-1)$ from Lemma~\ref{lemma:properties of D}\ref{(i)}.  Therefore $\hat{\text{df}}_2 = \mathrm{tr}(\mathbf{P})= p$.

\begin{figure}[htp]
\begin{center}
\includegraphics[scale=0.57]{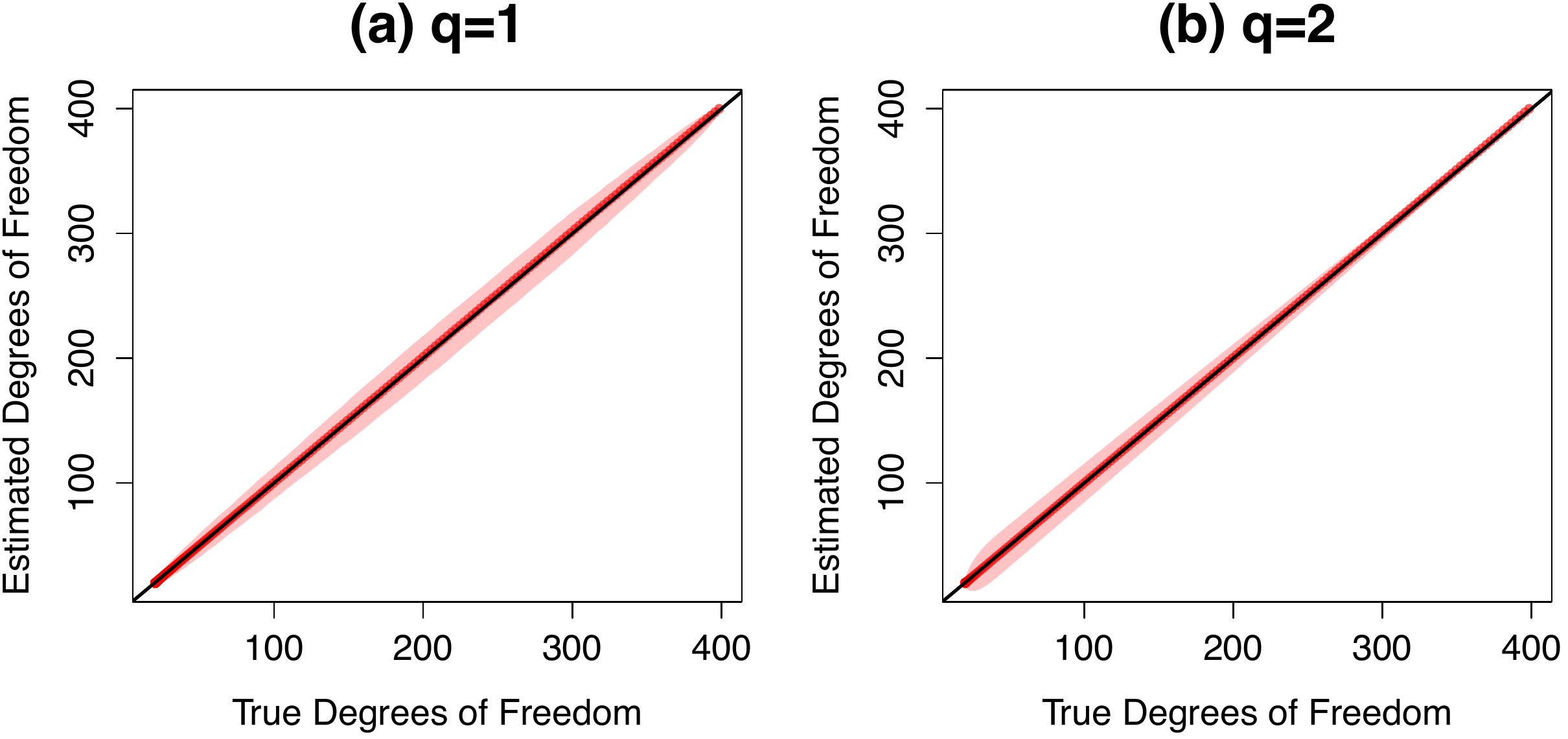}
\end{center}
\caption{\label{Fig:dofl1} We compare the true degrees of freedom of convex clustering ($x$-axis), given in (\ref{Eq:montecarlo dof}), to the proposed unbiased estimators of the degrees of freedom ($y$-axis), given in Lemmas~\ref{dof:l1} and \ref{lemma:dof:l2}.  Panels (a) and (b) contain the results for convex clustering with $q=1$ and $q=2$, respectively.   The red line is the mean of the estimated degrees of freedom for convex clustering over 500 data sets, obtained by varying the tuning parameter $\lambda$. The 
shaded bands indicate the point-wise intervals of the estimated degrees of freedom (mean $\pm$ 2 $\times$ standard deviation), over 500  data sets.    The black line indicates $y=x$.  }
\end{figure}

We now assess the accuracy of the proposed unbiased estimators of the degrees of freedom.  We simulate Gaussian clusters with $K=2$ as described in Section~\ref{subsec:gaussian clusters} with $n=p=20$ and $\sigma=0.5$.  We perform convex clustering with $q=1$ and $q=2$ across a fine grid of tuning parameters $\lambda$.   For each $\lambda$, we compare the quantities (\ref{Eq:dof1}) and~(\ref{Eq:dof2}) to
\begin{equation}
\label{Eq:montecarlo dof}
\frac{1}{\sigma^2} \sum_{j=1}^{np} (\hat{u}_j-u_j)(x_j-u_j),
\end{equation}
 which is an unbiased estimator of the true degrees of freedom, $\frac{1}{\sigma^2} \sum_{j=1}^{np} \mathrm{Cov}(\hat{u}_j,x_j)$, averaged over 500 data sets.  In addition, we plot the point-wise intervals of the estimated degrees of freedom (mean $\pm$ 2 $\times$ standard deviation).  Note that (\ref{Eq:montecarlo dof}) cannot be computed in practice, since it requires knowledge of the unknown quantity $\mathbf{u}$.  Results are displayed in Figure~\ref{Fig:dofl1}. We see that the estimated degrees of freedom are quite close to the true degrees of freedom.  
 
\section{Simulation Studies}
\label{Sec:sim study}
We compare convex clustering with $q=1$ and $q=2$ to the following proposals:
\begin{enumerate}
\item Single linkage hierarchical clustering with the dissimilarity matrix defined by the Euclidean distance between two observations.  
\item The $k$-means clustering algorithm \citep{lloyd1982least}.
\item Average linkage hierarchical clustering with the dissimilarity matrix defined by the Euclidean distance between two observations.
\end{enumerate}

We implement convex clustering (\ref{Eq:convex clustering2}) with $q=\{1,2\}$ using the \verb=R= package \verb=cvxclustr= \citep{cvxclustr}.  In order to obtain the entire solution path for convex clustering,  we use a fine grid of $\lambda$ values for (\ref{Eq:convex clustering2}), in a range guided by Lemma~\ref{lemma:max lambda}.  We apply the other methods by allowing the number of clusters to vary over a range from $1$ to $n$ clusters.
To evaluate and quantify the performance of the different clustering methods, we use the Rand index \citep{rand1971objective}.  A high value of the Rand index indicates good agreement between the true and estimated clusters.

We consider two different types of clusters in our simulation studies: Gaussian clusters and non-convex clusters.

\subsection{Gaussian Clusters}
\label{subsec:gaussian clusters}
\begin{figure}[htp]
\begin{center}
\includegraphics[scale=0.475]{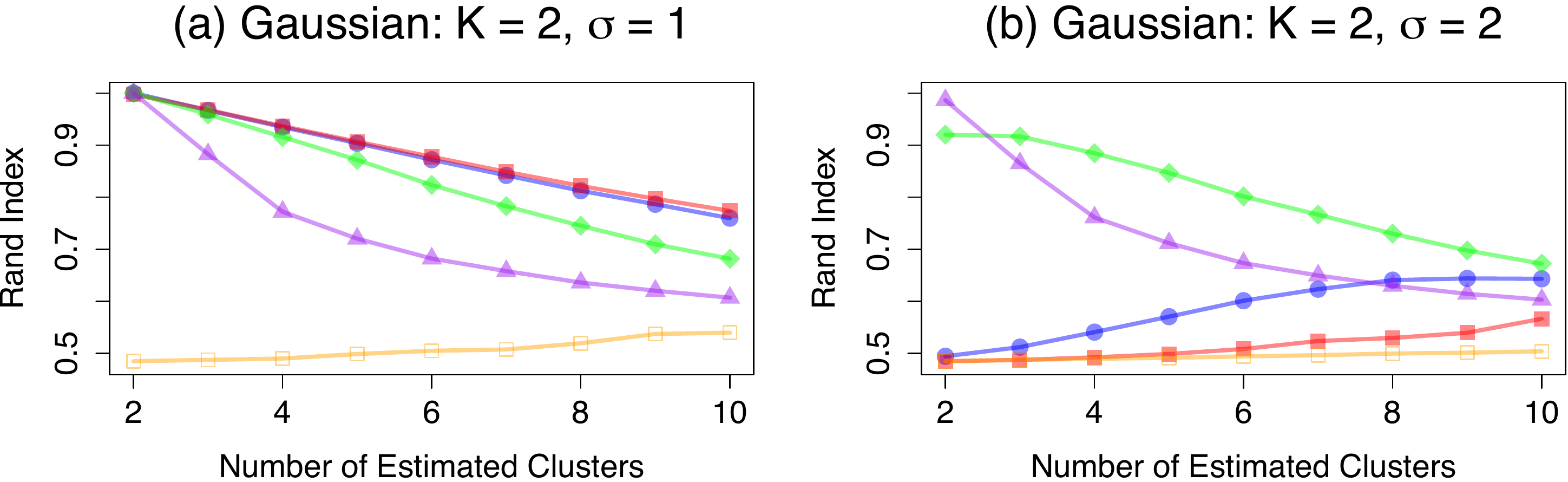}
\end{center}
\caption{\label{Fig:convexK2} Simulation results for Gaussian clusters with $K=2$, $n=p=30$, averaged over 200 data sets, for two noise levels $\sigma=\{1,2\}$.  Colored lines correspond to  single linkage hierarchical 
clustering (\protect\includegraphics[height=0.4em]{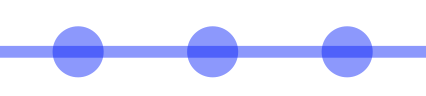}), average linkage hierarchical clustering (\protect\includegraphics[height=0.4em]{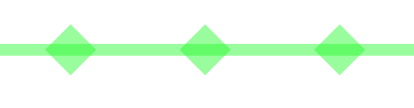}), $k$-means clustering (\protect\includegraphics[height=0.4em]{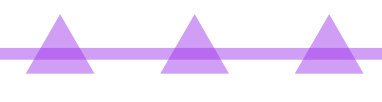}), convex clustering with $q=1$  (\protect\includegraphics[height=0.4em]{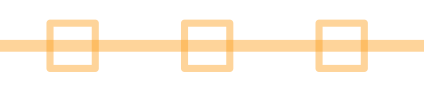}), and convex clustering with $q=2$  (\protect\includegraphics[height=0.4em]{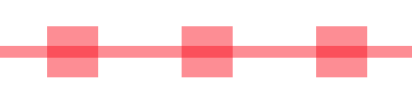}).      }
\end{figure}

\begin{figure}[htp]
\begin{center}
\includegraphics[scale=0.475]{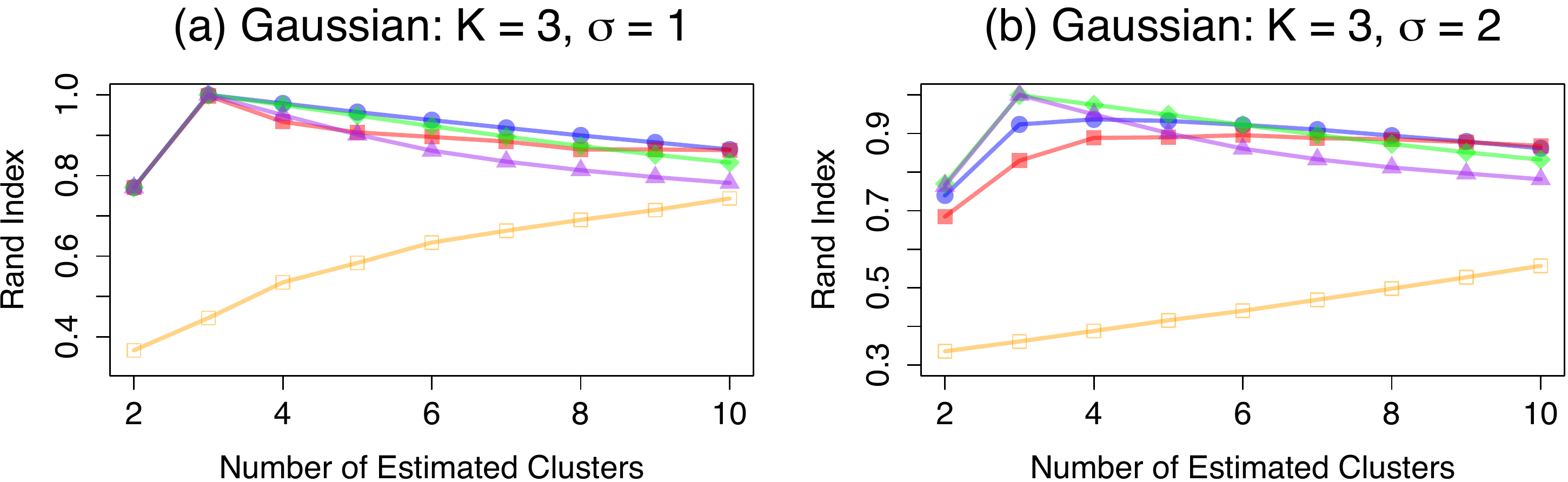}
\end{center}
\caption{\label{Fig:convexK3}  Simulation results for Gaussian clusters with $K=3$, $n=p=30$, averaged over 200 data sets, for two noise levels $\sigma=\{1,2\}$.  Line types are as described in Figure~\ref{Fig:convexK2}.       }
\end{figure}

 We generate Gaussian clusters with $K=2$ and $K=3$ by randomly assigning each observation to a cluster with equal probability.  For $K=2$, we create the mean vectors $\boldsymbol{\mu}_1 = \mathbf{1}_p$ and $\boldsymbol{\mu}_2= -\mathbf{1}_p$.  For $K=3$, we create the mean vectors $\boldsymbol{\mu}_{1} = -3 \cdot \mathbf{1}_p$, $\boldsymbol{\mu}_{2} = \mathbf{0}_p$, and $\boldsymbol{\mu}_{3} = 3 \cdot \mathbf{1}_p$.
 We then generate the $n\times p$ data matrix 
 $\mathbf{X}$ according to $\mathbf{X}_{i.} \sim \mathrm{MVN}(\boldsymbol{\mu}_k,\sigma^2 \mathbf{I})$ for $i\in D_k$.  We consider $n=p=30$ and $\sigma = \{1,2\}$.
The Rand indices for $K=2$ and $K=3$, averaged over 200 data sets, are summarized in Figures~\ref{Fig:convexK2} and~\ref{Fig:convexK3}, respectively.

Recall from Section~\ref{subsec:connection slc} that there is a connection between convex clustering and single linkage clustering.   
However, we note that the two clustering methods are not equivalent. From Figure~\ref{Fig:convexK2}(a), we see that single linkage hierarchical clustering performs very similarly to convex clustering with $q=2$ when the signal-to-noise ratio is high.  However, from Figure~\ref{Fig:convexK2}(b), we see that single linkage hierarchical clustering outperforms convex clustering with $q=2$   when the signal-to-noise ratio is low.

We also established a connection between convex clustering and $k$-means clustering in Section~\ref{subsec:kmeans}.  From Figure~\ref{Fig:convexK2}(a), we see that $k$-means clustering and convex clustering with $q=2$ perform similarly when two clusters are estimated and the signal-to-noise ratio is high, since in this case the penalty term dominates the first term in~(\ref{Eq:cc-kmeans3}).   In contrast, when the signal-to-noise ratio is low, the first term dominates the penalty term in~(\ref{Eq:cc-kmeans3}).  Therefore, when convex clustering with $q=2$ estimates two clusters, one cluster is of size one and the other is of size $n-1$, as discussed in Section~\ref{subsec:kmeans}.  
Figure~\ref{Fig:convexK2}(b) illustrates this phenomenon when both methods estimate two clusters: convex clustering with $q=2$ has a Rand index of approximately 0.5 while $k$-means clustering has a Rand index of one.
    
 All methods outperform convex clustering with $q=1$.  Moreover, $k$-means clustering and average linkage hierarchical clustering outperform single linkage hierarchical clustering and convex clustering when the signal-to-noise ratio is low. This suggests that the minimum signal needed for convex clustering to identify the correct clusters may be larger than that of average linkage hierarchical clustering and $k$-means clustering.  
We see similar results for the case when $K=3$ in Figure~\ref{Fig:convexK3}.

\subsection{Non-Convex Clusters}
\label{subsec:nonconvex}
\begin{figure}[htp]
\begin{center}
\includegraphics[scale=0.475]{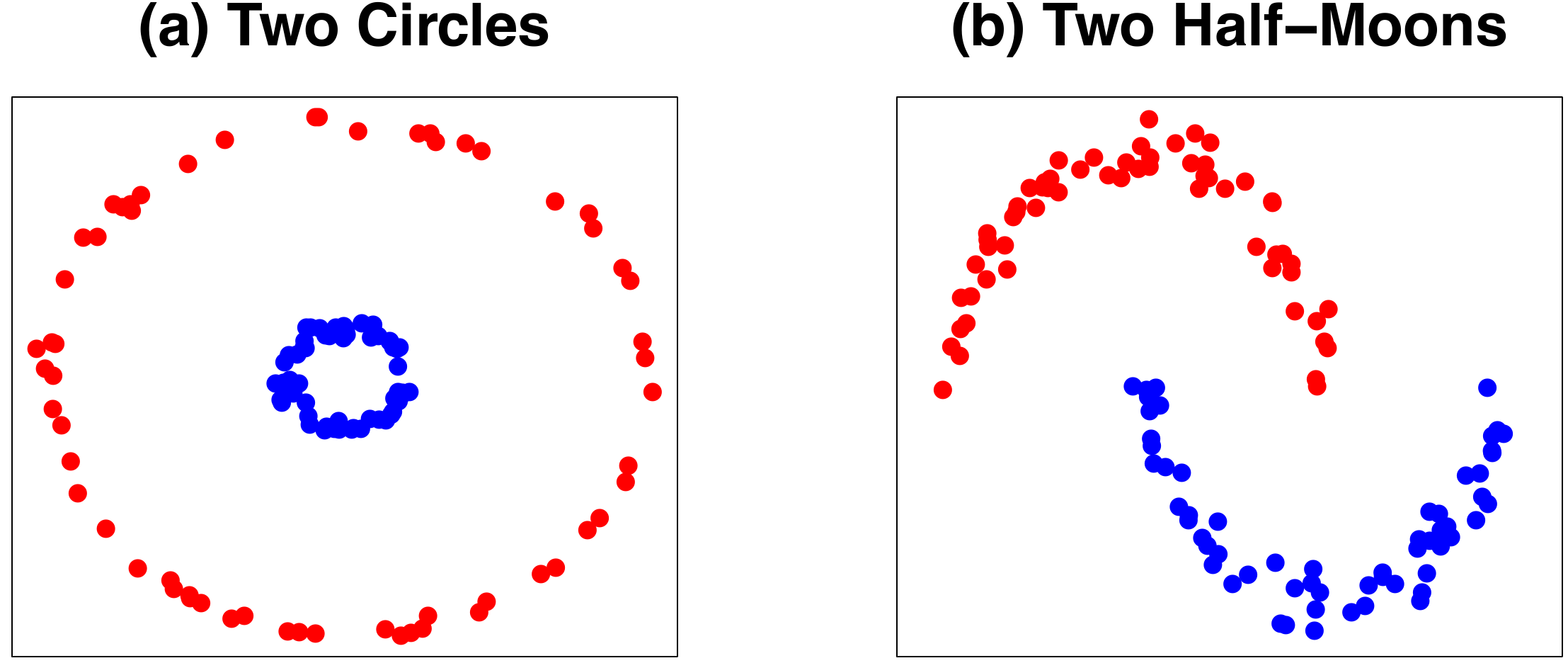}
\end{center}
\caption{\label{Fig:nonconvextoy}   Illustrations of two circles clusters and two half-moons clusters with $n=100$.   }
\end{figure}

\begin{figure}[htp]
\begin{center}
\includegraphics[scale=0.475]{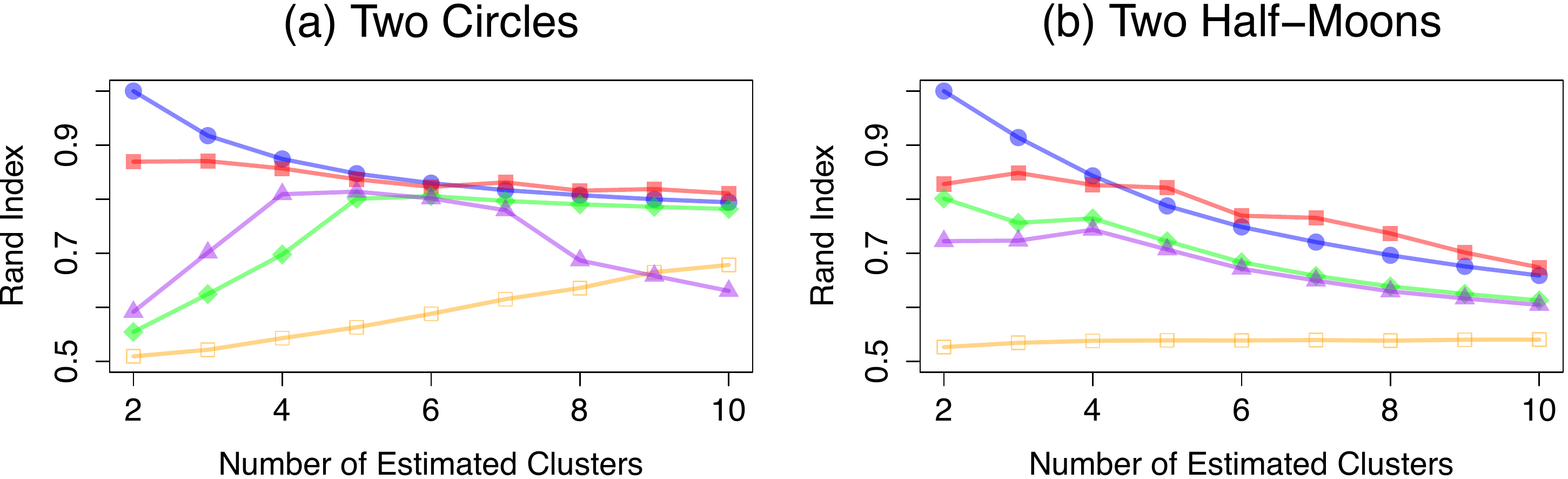}
\end{center}
\caption{\label{Fig:nonconvex}  Simulation results for the two circles and two half-moons clusters   
with $n=100$, averaged over 200 data sets.  Line types are as described in Figure~\ref{Fig:convexK2}.      }
\end{figure}

We consider two types of non-convex clusters: two \emph{circles clusters} \citep{ng2002spectral} and two \emph{half-moon clusters} \citep{hocking2011clusterpath,chi2013splitting}.  For two circles clusters, we generate 50 data points from each of the two circles that are centered at $(0,0)$ with radiuses two and 10, respectively.  We then add Gaussian random noise with mean zero and standard deviation 0.1 to each data point. For two half-moon clusters, we generate 50 data points from each of the two half-circles that are centered at $(0,0)$ and $(30,3)$ with radius 30, respectively.   We then add Gaussian random noise with mean zero and standard deviation one to each data point.    
Illustrations of both types of clusters are given in Figure~\ref{Fig:nonconvextoy}.
The Rand indices for both types of clusters, averaged over 200 data sets, are summarized in Figure~\ref{Fig:nonconvex}.

We see from Figure~\ref{Fig:nonconvex} that convex clustering with $q=2$ and single linkage hierarchical clustering have similar performance, and that they outperform all of the other methods. Single linkage hierarchical clustering is able to identify non-convex clusters since it is an agglomerative algorithm that merges the closest pair of observations not yet belonging to the same cluster into one cluster.  
In contrast, average linkage hierarchical clustering and $k$-means clustering are known to perform poorly on identifying non-convex clusters \citep{ng2002spectral,hocking2011clusterpath}.  Again, convex clustering with $q=1$ has the worst performance.

\subsection{Selection of the Tuning Parameter $\lambda$}
\label{Sec:selection of lambda}
Convex clustering~(\ref{Eq:convex clustering2}) involves a tuning parameter $\lambda$, which determines the estimated number of clusters.  
Some authors have suggested a hold-out validation approach to select   tuning parameters for clustering problems (see, for instance, \citealp{tan2014sparse,chi2014convex}).
In this section, we present an alternative approach for selecting $\lambda$ using the unbiased estimators of the degrees of freedom derived in Section~\ref{Sec:degrees of freedom}.

The Bayesian Information Criterion (BIC) developed in \citet{schwarz1978estimating} has been used extensively for model selection.  
However, it is known that the BIC does not perform well unless the number of observations is far larger than the number of parameters \citep{chen2008extended,chen2012extended}.  
For convex clustering~(\ref{Eq:convex clustering2}), the number of observations is equal to the number of parameters.  
Thus, we consider the extended BIC \citep{chen2008extended,chen2012extended}, defined as
\begin{equation}
\label{Eq:eBIC}
\mathrm{eBIC}_{q,\gamma} = np\cdot \log \left(\frac{\mathrm{RSS}_{q}}{np} \right) + \hat{\mathrm{df}}_{q} \cdot \log (np) + 2\gamma \cdot \hat{\mathrm{df}}_{q}\cdot \log (np),
\end{equation}
where $\mathrm{RSS}_{q} = \|\mathbf{x}-\hat{\mathbf{u}}_{q}\|_2^2$, $\hat{\mathbf{u}}_{q}$ is the convex clustering estimate for a given value of $q$ and $\lambda$, $\gamma \in [0,1]$, and $\hat{\mathrm{df}}_q$ is given in Section~\ref{Sec:degrees of freedom}.
Note that we suppress the dependence of $\hat{\mathbf{u}}_q$ and $\hat{\mathrm{df}}_{q}$ on $\lambda$ for notational convenience.
We see that when $\gamma = 0$, the extended BIC reduces to the classical BIC. 

To evaluate the performance of the extended BIC in selecting the number of clusters, we generate Gaussian clusters with $K=2$ and $K=3$ as described in Section~\ref{subsec:gaussian clusters}, with $n=p=20$, and $\sigma=0.5$.  We perform convex clustering with $q=2$ over a fine grid of $\lambda$, and select the value of $\lambda$ for which the quantity $\mathrm{eBIC}_{q,\gamma}$ is minimized. We consider $\gamma\in \{0,0.5,0.75,1\}$.  Table~\ref{Table:inf1} 
reports the proportion of datasets for which the correct number of clusters was identified, as well as the average Rand index.

From Table~\ref{Table:inf1}, we see that the extended BIC is able to select the true number of clusters accurately for $K=2$.  When $K=3$, the classical BIC ($\gamma =0$) fails to select the true number of clusters. In contrast, the extended BIC with $\gamma=1$ has the best performance.

\begin{table}[htp]
\footnotesize
\begin{center}
\caption{Simulation study to evaluate the performance of the extended BIC for tuning parameter selection for convex clustering with $q=2$.  Results are reported over 100 simulated data sets.  We report the proportion of data sets for which the correct number of clusters was identified, and the average Rand index.  }
\begin{tabular}{|c|l|   c c|}
  \hline
&$\mathrm{eBIC}_{2,\gamma}$&Correct number of clusters &Rand index \\ \hline
Gaussian clusters, $K=2$&$\gamma = 0$ & 0.94 &  0.9896\\
&$\gamma = 0.5$ & 0.98 &  0.9991 \\
&$\gamma = 0.75$ & 0.99 &  0.9995 \\
&$\gamma = 1$ & 0.99 &  0.9995  \\\hline
Gaussian clusters, $K=3$&$\gamma = 0$ & 0.06 &  0.7616\\
&$\gamma = 0.5$ & 0.59 &  0.9681 \\
&$\gamma = 0.75$ & 0.70 &  0.9768 \\
&$\gamma = 1$ & 0.84 &  0.9873  \\\hline

\end{tabular}
\label{Table:inf1}
\end{center}
\end{table}

\section{Discussion}
\label{Sec:discussion}
Convex clustering recasts the clustering problem into a penalized regression problem.  By studying its dual problem, we show that there is a connection between convex clustering and single linkage hierarchical clustering. In addition, we establish  a connection between convex clustering and $k$-means clustering.  We also establish several statistical properties of convex clustering.  Through some numerical studies, we illustrate that the performance of convex clustering may not be appealing relative to traditional  clustering methods, especially when the signal-to-noise ratio is low.

Many authors have proposed a modification to the convex clustering problem (\ref{Eq:convex clustering}),
\begin{equation}
\label{Eq:convex clustering weights}
\underset{\mathbf{U}\in \mathbb{R}^{n\times p}}{\mathrm{minimize}} \; \frac{1}{2}\sum_{i=1}^n \|\mathbf{X}_{i.}- \mathbf{U}_{i.}\|^2_2 + \lambda \mathrm{Q}_q (\mathbf{W},\mathbf{U}),
\end{equation}  
where $\mathbf{W}$ is an $n\times n$ symmetric  matrix of positive weights, and $\mathrm{Q}_q  (\mathbf{W},\mathbf{U}) = \sum_{i<i'} W_{ii'}\|\mathbf{U}_{i.}-\mathbf{U}_{i'.}\|_q$ \citep{pelckmans2005convex,hocking2011clusterpath,lindsten2011clustering,chi2013splitting}. For instance, the weights can be defined as $W_{ii'} = \exp \left( - \phi \|\mathbf{X}_{i.}-\mathbf{X}_{i'.}\|_2^2 \right)$ for some constant $\phi >0$.  This yields better empirical performance than (\ref{Eq:convex clustering}) \citep{hocking2011clusterpath,chi2013splitting}.
We leave an investigation of the properties of (\ref{Eq:convex clustering weights}) to future work.

\section*{Acknowledgment} We thank Ashley Petersen, Ali Shojaie, and Noah Simon for helpful conversations on earlier drafts of this manuscript. We thank the editor and two reviewers for helpful comments that improved the quality of this manuscript.  D. W. was partially supported by a Sloan Research Fellowship, NIH Grant DP5OD009145, and NSF CAREER DMS-1252624.

\bibliography{reference}

\begin{thebibliography}{30}

\bibitem[\protect\citeauthoryear{Bach et~al.}{2011}]{bach2011convex}
\begin{barticle}[author]
\bauthor{\bsnm{Bach},~\bfnm{Francis}\binits{F.}},
  \bauthor{\bsnm{Jenatton},~\bfnm{Rodolphe}\binits{R.}},
  \bauthor{\bsnm{Mairal},~\bfnm{Julien}\binits{J.}} \AND
  \bauthor{\bsnm{Obozinski},~\bfnm{Guillaume}\binits{G.}}
(\byear{2011}).
\btitle{Convex optimization with sparsity-inducing norms}.
\bjournal{Optimization for Machine Learning}
\bpages{19--53}.
\end{barticle}
\endbibitem

\bibitem[\protect\citeauthoryear{Boucheron, Lugosi and
  Massart}{2013}]{boucheron2013concentration}
\begin{bbook}[author]
\bauthor{\bsnm{Boucheron},~\bfnm{St{\'e}phane}\binits{S.}},
  \bauthor{\bsnm{Lugosi},~\bfnm{G{\'a}bor}\binits{G.}} \AND
  \bauthor{\bsnm{Massart},~\bfnm{Pascal}\binits{P.}}
(\byear{2013}).
\btitle{Concentration inequalities: A nonasymptotic theory of independence}.
\bpublisher{OUP Oxford}.
\end{bbook}
\endbibitem

\bibitem[\protect\citeauthoryear{Boyd and Vandenberghe}{2004}]{boyd2004convex}
\begin{bbook}[author]
\bauthor{\bsnm{Boyd},~\bfnm{Stephen}\binits{S.}} \AND
  \bauthor{\bsnm{Vandenberghe},~\bfnm{Lieven}\binits{L.}}
(\byear{2004}).
\btitle{Convex Optimization}.
\bpublisher{Cambridge university press}.
\end{bbook}
\endbibitem

\bibitem[\protect\citeauthoryear{Chen and Chen}{2008}]{chen2008extended}
\begin{barticle}[author]
\bauthor{\bsnm{Chen},~\bfnm{Jiahua}\binits{J.}} \AND
  \bauthor{\bsnm{Chen},~\bfnm{Zehua}\binits{Z.}}
(\byear{2008}).
\btitle{Extended {Bayesian} information criteria for model selection with large
  model spaces}.
\bjournal{Biometrika}
\bvolume{95}
\bpages{759--771}.
\end{barticle}
\endbibitem

\bibitem[\protect\citeauthoryear{Chen and Chen}{2012}]{chen2012extended}
\begin{barticle}[author]
\bauthor{\bsnm{Chen},~\bfnm{Jiahua}\binits{J.}} \AND
  \bauthor{\bsnm{Chen},~\bfnm{Zehua}\binits{Z.}}
(\byear{2012}).
\btitle{{Extended BIC for small-$n$-large-$P$ sparse GLM}}.
\bjournal{Statistica Sinica}
\bvolume{22}
\bpages{555}.
\end{barticle}
\endbibitem

\bibitem[\protect\citeauthoryear{Chi, Allen and Baraniuk}{2014}]{chi2014convex}
\begin{barticle}[author]
\bauthor{\bsnm{Chi},~\bfnm{Eric~C}\binits{E.~C.}},
  \bauthor{\bsnm{Allen},~\bfnm{Genevera~I}\binits{G.~I.}} \AND
  \bauthor{\bsnm{Baraniuk},~\bfnm{Richard~G}\binits{R.~G.}}
(\byear{2014}).
\btitle{Convex Biclustering}.
\bjournal{arXiv preprint arXiv:1408.0856}.
\end{barticle}
\endbibitem

\bibitem[\protect\citeauthoryear{Chi and Lange}{2014a}]{chi2013splitting}
\begin{barticle}[author]
\bauthor{\bsnm{Chi},~\bfnm{Eric}\binits{E.}} \AND
  \bauthor{\bsnm{Lange},~\bfnm{Kenneth}\binits{K.}}
(\byear{2014}a).
\btitle{Splitting methods for convex clustering}.
\bjournal{Journal of Computational and Graphical Statistics, in press}.
\end{barticle}
\endbibitem

\bibitem[\protect\citeauthoryear{Chi and Lange}{2014b}]{cvxclustr}
\begin{bmanual}[author]
\bauthor{\bsnm{Chi},~\bfnm{Eric}\binits{E.}} \AND
  \bauthor{\bsnm{Lange},~\bfnm{Kenneth}\binits{K.}}
(\byear{2014}b).
\btitle{cvxclustr: Splitting methods for convex clustering}
\bnote{URL \tt{http://cran.r-project.org/web/packages/cvxclustr.} \tt{R}
  package version 1.1.1}.
\end{bmanual}
\endbibitem

\bibitem[\protect\citeauthoryear{Duchi and Singer}{2009}]{duchisinger2009}
\begin{barticle}[author]
\bauthor{\bsnm{Duchi},~\bfnm{John}\binits{J.}} \AND
  \bauthor{\bsnm{Singer},~\bfnm{Yoram}\binits{Y.}}
(\byear{2009}).
\btitle{Efficient online and batch learning using forward backward splitting}.
\bjournal{The Journal of Machine Learning Research}
\bvolume{10}
\bpages{2899--2934}.
\end{barticle}
\endbibitem

\bibitem[\protect\citeauthoryear{Efron}{1986}]{efron1986biased}
\begin{barticle}[author]
\bauthor{\bsnm{Efron},~\bfnm{Bradley}\binits{B.}}
(\byear{1986}).
\btitle{How biased is the apparent error rate of a prediction rule?}
\bjournal{Journal of the American Statistical Association}
\bvolume{81}
\bpages{461--470}.
\end{barticle}
\endbibitem

\bibitem[\protect\citeauthoryear{Hanson and Wright}{1971}]{hanson1971bound}
\begin{barticle}[author]
\bauthor{\bsnm{Hanson},~\bfnm{David~Lee}\binits{D.~L.}} \AND
  \bauthor{\bsnm{Wright},~\bfnm{Farroll~Tim}\binits{F.~T.}}
(\byear{1971}).
\btitle{A bound on tail probabilities for quadratic forms in independent random
  variables}.
\bjournal{The Annals of Mathematical Statistics}
\bvolume{42}
\bpages{1079--1083}.
\end{barticle}
\endbibitem

\bibitem[\protect\citeauthoryear{Haris, Witten and
  Simon}{2015}]{haris2014convex}
\begin{barticle}[author]
\bauthor{\bsnm{Haris},~\bfnm{Asad}\binits{A.}},
  \bauthor{\bsnm{Witten},~\bfnm{Daniela}\binits{D.}} \AND
  \bauthor{\bsnm{Simon},~\bfnm{Noah}\binits{N.}}
(\byear{2015}).
\btitle{Convex modeling of interactions with strong heredity}.
\bjournal{Journal of Computational and Graphical Statistics, in press}.
\end{barticle}
\endbibitem

\bibitem[\protect\citeauthoryear{Hastie, Tibshirani and
  Friedman}{2009}]{ElemStatLearn}
\begin{bbook}[author]
\bauthor{\bsnm{Hastie},~\bfnm{T.}\binits{T.}},
  \bauthor{\bsnm{Tibshirani},~\bfnm{R.}\binits{R.}} \AND
  \bauthor{\bsnm{Friedman},~\bfnm{J.}\binits{J.}}
(\byear{2009}).
\btitle{The Elements of Statistical Learning; Data Mining, Inference and
  Prediction}.
\bpublisher{Springer Verlag}, \baddress{New York}.
\end{bbook}
\endbibitem

\bibitem[\protect\citeauthoryear{Hocking et~al.}{2011}]{hocking2011clusterpath}
\begin{binproceedings}[author]
\bauthor{\bsnm{Hocking},~\bfnm{Toby~Dylan}\binits{T.~D.}},
  \bauthor{\bsnm{Joulin},~\bfnm{Armand}\binits{A.}},
  \bauthor{\bsnm{Bach},~\bfnm{Francis}\binits{F.}},
  \bauthor{\bsnm{Vert},~\bfnm{Jean-Philippe}\binits{J.-P.}} \betal{et~al.}
(\byear{2011}).
\btitle{Clusterpath: an algorithm for clustering using convex fusion
  penalties}.
In \bbooktitle{28th International Conference on Machine Learning}.
\end{binproceedings}
\endbibitem

\bibitem[\protect\citeauthoryear{Jain and Dubes}{1988}]{JD88}
\begin{bbook}[author]
\bauthor{\bsnm{Jain},~\bfnm{A.~K.}\binits{A.~K.}} \AND
  \bauthor{\bsnm{Dubes},~\bfnm{R.~C.}\binits{R.~C.}}
(\byear{1988}).
\btitle{Algorithms for Clustering Data}.
\bpublisher{Prentice-Hall}.
\end{bbook}
\endbibitem

\bibitem[\protect\citeauthoryear{Lindsten, Ohlsson and
  Ljung}{2011}]{lindsten2011clustering}
\begin{binproceedings}[author]
\bauthor{\bsnm{Lindsten},~\bfnm{Fredrik}\binits{F.}},
  \bauthor{\bsnm{Ohlsson},~\bfnm{Henrik}\binits{H.}} \AND
  \bauthor{\bsnm{Ljung},~\bfnm{Lennart}\binits{L.}}
(\byear{2011}).
\btitle{Clustering using sum-of-norms regularization: With application to
  particle filter output computation}.
In \bbooktitle{Statistical Signal Processing Workshop (SSP)}
\bpages{201--204}.
\bpublisher{IEEE}.
\end{binproceedings}
\endbibitem

\bibitem[\protect\citeauthoryear{Liu, Yuan and Ye}{2013}]{liu2013guaranteed}
\begin{binproceedings}[author]
\bauthor{\bsnm{Liu},~\bfnm{Ji}\binits{J.}},
  \bauthor{\bsnm{Yuan},~\bfnm{Lei}\binits{L.}} \AND
  \bauthor{\bsnm{Ye},~\bfnm{Jieping}\binits{J.}}
(\byear{2013}).
\btitle{Guaranteed sparse recovery under linear transformation}.
In \bbooktitle{Proceedings of the 30th International Conference on Machine
  Learning (ICML-13)}
\bpages{91--99}.
\end{binproceedings}
\endbibitem

\bibitem[\protect\citeauthoryear{Lloyd}{1982}]{lloyd1982least}
\begin{barticle}[author]
\bauthor{\bsnm{Lloyd},~\bfnm{Stuart}\binits{S.}}
(\byear{1982}).
\btitle{{Least squares quantization in PCM}}.
\bjournal{IEEE Transactions on Information Theory}
\bvolume{28}
\bpages{129--137}.
\end{barticle}
\endbibitem

\bibitem[\protect\citeauthoryear{Ng, Jordan and Weiss}{2002}]{ng2002spectral}
\begin{barticle}[author]
\bauthor{\bsnm{Ng},~\bfnm{Andrew~Y}\binits{A.~Y.}},
  \bauthor{\bsnm{Jordan},~\bfnm{Michael~I}\binits{M.~I.}} \AND
  \bauthor{\bsnm{Weiss},~\bfnm{Yair}\binits{Y.}}
(\byear{2002}).
\btitle{On spectral clustering: Analysis and an algorithm}.
\bjournal{Advances in Neural Information Processing Systems}.
\end{barticle}
\endbibitem

\bibitem[\protect\citeauthoryear{Pelckmans et~al.}{2005}]{pelckmans2005convex}
\begin{binproceedings}[author]
\bauthor{\bsnm{Pelckmans},~\bfnm{Kristiaan}\binits{K.}},
  \bauthor{\bsnm{De~Brabanter},~\bfnm{Joseph}\binits{J.}},
  \bauthor{\bsnm{Suykens},~\bfnm{JAK}\binits{J.}} \AND
  \bauthor{\bsnm{De~Moor},~\bfnm{B}\binits{B.}}
(\byear{2005}).
\btitle{Convex clustering shrinkage}.
In \bbooktitle{PASCAL Workshop on Statistics and Optimization of Clustering
  Workshop}.
\end{binproceedings}
\endbibitem

\bibitem[\protect\citeauthoryear{Radchenko and
  Mukherjee}{2014}]{radchenko2014consistent}
\begin{barticle}[author]
\bauthor{\bsnm{Radchenko},~\bfnm{Peter}\binits{P.}} \AND
  \bauthor{\bsnm{Mukherjee},~\bfnm{Gourab}\binits{G.}}
(\byear{2014}).
\btitle{Consistent clustering using $\ell_1$ fusion penalty}.
\bjournal{arXiv preprint arXiv:1412.0753}.
\end{barticle}
\endbibitem

\bibitem[\protect\citeauthoryear{Rand}{1971}]{rand1971objective}
\begin{barticle}[author]
\bauthor{\bsnm{Rand},~\bfnm{William~M}\binits{W.~M.}}
(\byear{1971}).
\btitle{Objective criteria for the evaluation of clustering methods}.
\bjournal{Journal of the American Statistical association}
\bvolume{66}
\bpages{846--850}.
\end{barticle}
\endbibitem

\bibitem[\protect\citeauthoryear{Schwarz}{1978}]{schwarz1978estimating}
\begin{barticle}[author]
\bauthor{\bsnm{Schwarz},~\bfnm{Gideon}\binits{G.}}
(\byear{1978}).
\btitle{Estimating the dimension of a model}.
\bjournal{The Annals of Statistics}
\bvolume{6}
\bpages{461--464}.
\end{barticle}
\endbibitem

\bibitem[\protect\citeauthoryear{Tan and Witten}{2014}]{tan2014sparse}
\begin{barticle}[author]
\bauthor{\bsnm{Tan},~\bfnm{Kean~Ming}\binits{K.~M.}} \AND
  \bauthor{\bsnm{Witten},~\bfnm{Daniela~M}\binits{D.~M.}}
(\byear{2014}).
\btitle{Sparse biclustering of transposable data}.
\bjournal{Journal of Computational and Graphical Statistics}
\bvolume{23}
\bpages{985--1008}.
\end{barticle}
\endbibitem

\bibitem[\protect\citeauthoryear{Tibshirani and
  Taylor}{2011}]{tibshirani2011solution}
\begin{barticle}[author]
\bauthor{\bsnm{Tibshirani},~\bfnm{Ryan~J}\binits{R.~J.}} \AND
  \bauthor{\bsnm{Taylor},~\bfnm{Jonathan}\binits{J.}}
(\byear{2011}).
\btitle{The solution path of the generalized lasso}.
\bjournal{The Annals of Statistics}
\bvolume{39}
\bpages{1335--1371}.
\end{barticle}
\endbibitem

\bibitem[\protect\citeauthoryear{Tibshirani and
  Taylor}{2012}]{tibshirani2012degrees}
\begin{barticle}[author]
\bauthor{\bsnm{Tibshirani},~\bfnm{Ryan~J}\binits{R.~J.}} \AND
  \bauthor{\bsnm{Taylor},~\bfnm{Jonathan}\binits{J.}}
(\byear{2012}).
\btitle{Degrees of freedom in lasso problems}.
\bjournal{The Annals of Statistics}
\bvolume{40}
\bpages{1198--1232}.
\end{barticle}
\endbibitem

\bibitem[\protect\citeauthoryear{Tibshirani
  et~al.}{2005}]{tibshirani2005sparsity}
\begin{barticle}[author]
\bauthor{\bsnm{Tibshirani},~\bfnm{Robert}\binits{R.}},
  \bauthor{\bsnm{Saunders},~\bfnm{Michael}\binits{M.}},
  \bauthor{\bsnm{Rosset},~\bfnm{Saharon}\binits{S.}},
  \bauthor{\bsnm{Zhu},~\bfnm{Ji}\binits{J.}} \AND
  \bauthor{\bsnm{Knight},~\bfnm{Keith}\binits{K.}}
(\byear{2005}).
\btitle{Sparsity and smoothness via the fused lasso}.
\bjournal{Journal of the Royal Statistical Society: Series B (Statistical
  Methodology)}
\bvolume{67}
\bpages{91--108}.
\end{barticle}
\endbibitem

\bibitem[\protect\citeauthoryear{Vaiter et~al.}{2014}]{vaiter2014degrees}
\begin{barticle}[author]
\bauthor{\bsnm{Vaiter},~\bfnm{Samuel}\binits{S.}},
  \bauthor{\bsnm{Deledalle},~\bfnm{Charles-Alban}\binits{C.-A.}},
  \bauthor{\bsnm{Peyr{\'e}},~\bfnm{Gabriel}\binits{G.}},
  \bauthor{\bsnm{Fadili},~\bfnm{Jalal~M}\binits{J.~M.}} \AND
  \bauthor{\bsnm{Dossal},~\bfnm{Charles}\binits{C.}}
(\byear{2014}).
\btitle{The degrees of freedom of partly smooth regularizers}.
\bjournal{arXiv preprint arXiv:1404.5557}.
\end{barticle}
\endbibitem

\bibitem[\protect\citeauthoryear{Witten and
  Tibshirani}{2010}]{witten2010framework}
\begin{barticle}[author]
\bauthor{\bsnm{Witten},~\bfnm{Daniela~M}\binits{D.~M.}} \AND
  \bauthor{\bsnm{Tibshirani},~\bfnm{Robert}\binits{R.}}
(\byear{2010}).
\btitle{A framework for feature selection in clustering}.
\bjournal{Journal of the American Statistical Association}
\bvolume{105}
\bpages{713--726}.
\end{barticle}
\endbibitem

\bibitem[\protect\citeauthoryear{Zhu et~al.}{2014}]{NIPS2014_5307}
\begin{bincollection}[author]
\bauthor{\bsnm{Zhu},~\bfnm{Changbo}\binits{C.}},
  \bauthor{\bsnm{Xu},~\bfnm{Huan}\binits{H.}},
  \bauthor{\bsnm{Leng},~\bfnm{Chenlei}\binits{C.}} \AND
  \bauthor{\bsnm{Yan},~\bfnm{Shuicheng}\binits{S.}}
(\byear{2014}).
\btitle{Convex optimization procedure for clustering: theoretical revisit}.
In \bbooktitle{Advances in Neural Information Processing Systems}.
\end{bincollection}
\endbibitem

\end{thebibliography}

\newpage
\appendix

\renewcommand{\theequation}{A-\arabic{equation}}
\setcounter{equation}{0}  
\section*{Appendix A: Proof of Lemmas~\ref{lemma:dual problem}---\ref{lemma:dual problem2}}  
\noindent{\textbf{Proof of Lemma~\ref{lemma:dual problem}:}
\begin{proof}
 We rewrite problem~(\ref{Eq:convex clustering2}) as 
\[
\underset{\mathbf{u}\in\mathbb{R}^{np}, \boldsymbol{\eta}_1\in \mathbb{R}^{\left[p\cdot {n\choose 2}\right]} }{\mathrm{minimize}} \; \frac{1}{2}\|\mathbf{x-u}\|_2^2 + \lambda \mathrm{P}_{q} ( \boldsymbol{\eta}_1) \quad \quad \text{subject to } \boldsymbol{\eta}_1=\mathbf{Du},
\]
with the Lagrangian function 
\begin{equation}
\label{Eq:Lagrangian}
\mathcal{L}(\mathbf{u},\boldsymbol{\eta}_1,\boldsymbol{\nu}) = \frac{1}{2}\|\mathbf{x}-\mathbf{u}\|_2^2 + \lambda \mathrm{P}_q(\boldsymbol{\eta}_1) + \boldsymbol{\nu}^T (\mathbf{Du}-\boldsymbol{\eta}_1),
\end{equation}
where $\boldsymbol{\nu}\in \mathbb{R}^{\left[p \cdot {n\choose 2}\right]}$ is the Lagrangian dual variable.  
In order to derive the dual problem, we need to minimize the Lagrangian function over the primal variables $\mathbf{u}$ and $\boldsymbol{\eta}_1$.  Recall from (\ref{dual norm})  that  $\mathrm{P}^*_q(\cdot)$ is the dual norm of $\mathrm{P}_q(\cdot)$. It can be shown that 
\[
\underset{\boldsymbol{\eta}_1\in \mathbb{R}^{\left[p\cdot {n \choose 2}\right]}}{\mathrm{inf}} \; \mathcal{L}(\mathbf{u},\boldsymbol{\eta}_1,\boldsymbol{\nu}) = \begin{cases} \frac{1}{2}\|\mathbf{x}-\mathbf{u}\|_2^2 + \boldsymbol{\nu}^T \mathbf{Du} & \text{if} \quad \mathrm{P}^*_q(\boldsymbol{\nu}) \le \lambda, \\ -\infty & \mathrm{otherwise,}\end{cases}
\]
and 
\[
\underset{\boldsymbol{\eta}_1\in \mathbb{R}^{\left[p\cdot {n \choose 2}\right]},\mathbf{u}\in \mathbb{R}^{np}}{\mathrm{inf}} \; \mathcal{L}(\mathbf{u},\boldsymbol{\eta}_1,\boldsymbol{\nu}) = \begin{cases} -\frac{1}{2}\|\mathbf{x}-\mathbf{D}^T\boldsymbol{\nu}\|_2^2 + \frac{1}{2}\|\mathbf{x}\|_2^2 & \text{if} \quad \mathrm{P}^*_q(\boldsymbol{\nu}) \le \lambda. \\ -\infty & \mathrm{otherwise.}\end{cases}
\]
Therefore, the dual problem for (\ref{Eq:convex clustering2}) is 
\begin{equation}
\label{Eq:dual problem for convex clustering2}
\underset{\boldsymbol{\nu}\in\mathbb{R}^{\left[p \cdot {n\choose 2}\right]}}{\mathrm{minimize}} \; \frac{1}{2}\|\mathbf{x}-\mathbf{D}^T\boldsymbol{\nu}\|_2^2 \quad \quad \mathrm{subject \; to}\; \mathrm{P}^*_q(\boldsymbol{\nu}) \le \lambda.
\end{equation}

We now establish an explicit relationship between the solution to convex clustering and its dual problem.  Differentiating the Lagrangian function~(\ref{Eq:Lagrangian}) with respect to $\mathbf{u}$ and setting it equal to zero, we obtain 
\[
 \hat{\mathbf{u}} = \mathbf{x}- \mathbf{D}^T\hat{\boldsymbol{\nu}},
\]
where $\hat{\boldsymbol{\nu}}$ is the solution to the dual problem, which satisfies $\mathrm{P}_q^*(\hat{\boldsymbol{\nu}})\le \lambda$ by (\ref{Eq:dual problem for convex clustering2}).
Multiplying both sides by $\mathbf{D}$, we obtain the  relationship~(\ref{Eq:convex clustering2 solution}).
\end{proof}

\noindent{\textbf{Proof of Lemma~\ref{lemma:dual problem2}:}
\begin{proof}
  We rewrite (\ref{Eq:convex modification}) as
\[
\underset{\boldsymbol{\gamma}\in \mathbb{R}^{\left[p\cdot {n \choose 2}\right]}, \boldsymbol{\eta}_2 \in \mathbb{R}^{\left[p \cdot {n\choose 2}\right]} }{\mathrm{minimize}}\; \frac{1}{2}\|\mathbf{Dx}-\boldsymbol{\gamma}\|_2^2 + \lambda \mathrm{P}_{q} (\boldsymbol{\eta}_2) \qquad \text{subject to } \boldsymbol{\eta}_2= \boldsymbol{\gamma},
\]
with the Lagrangian function 
\begin{equation}
\label{Eq:Lagrangian modification}
\mathcal{L}(\boldsymbol{\gamma},\boldsymbol{\eta}_2,\boldsymbol{\nu}') = \frac{1}{2}\|\mathbf{Dx}-\boldsymbol{\gamma}\|_2^2 + \lambda \mathrm{P}_q (\boldsymbol{\eta}_2) + (\boldsymbol{\nu}')^T (\boldsymbol{\gamma}-\boldsymbol{\eta}_2),
\end{equation}
where $\boldsymbol{\nu}' \in \mathbb{R}^{\left[ p \cdot {n\choose 2} \right]}$ is the Lagrangian dual variable.  In order to derive the dual problem, we minimize the Lagrangian function over the primal variables $\boldsymbol{\gamma}$ and $\boldsymbol{\eta}_2$.  It can be shown that 
\[
\underset{\boldsymbol{\eta}_2\in \mathbb{R}^{\left[p\cdot {n \choose 2}\right]}}{\mathrm{inf}} \;\mathcal{L}(\boldsymbol{\gamma},\boldsymbol{\eta}_2,\boldsymbol{\nu}') = \begin{cases} \frac{1}{2}\|\mathbf{Dx}-\boldsymbol{\gamma}\|_2^2 + (\boldsymbol{\nu}')^T \boldsymbol{\gamma} & \text{if} \quad \mathrm{P}^*_q(\boldsymbol{\nu}') \le \lambda, \\ -\infty & \mathrm{otherwise,}\end{cases}
\]
and 
\[
\underset{\boldsymbol{\eta}_2\in \mathbb{R}^{\left[p\cdot {n \choose 2}\right]},\boldsymbol{\gamma} \in \mathbb{R}^{\left[p\cdot {n \choose 2}\right]}}{\mathrm{inf}} \;\mathcal{L}(\boldsymbol{\gamma},\boldsymbol{\eta}_2,\boldsymbol{\nu}')= \begin{cases} -\frac{1}{2}\|\mathbf{Dx}-\boldsymbol{\nu}'\|_2^2 + \frac{1}{2}\|\mathbf{Dx}\|_2^2 & \text{if} \quad \mathrm{P}^*_q(\boldsymbol{\nu}') \le \lambda. \\ -\infty & \mathrm{otherwise.}\end{cases}
\]

\noindent Therefore, the dual problem for (\ref{Eq:convex modification}) is 
\begin{equation}
\label{Eq:dual problem for convex modification}
\underset{\boldsymbol{\nu}'\in\mathbb{R}^{\left[p \cdot {n\choose 2}\right]}}{\mathrm{minimize}} \; \frac{1}{2}\|\mathbf{Dx}-\boldsymbol{\nu}'\|_2^2 \quad \quad \text{subject to } \mathrm{P}^*_q(\boldsymbol{\nu}') \le \lambda.
\end{equation}
We now establish an explicit relationship between the solution to (\ref{Eq:convex modification}) and its dual problem.  Differentiating the Lagrangian function~(\ref{Eq:Lagrangian modification}) with respect to $\boldsymbol{\gamma}$ and setting it equal to zero, we obtain 
\[
\hat{\boldsymbol{\gamma}} = \mathbf{Dx}- \hat{\boldsymbol{\nu}}',
\]
where $\hat{\boldsymbol{\nu}}'$ is the solution to the dual problem, which we know from (\ref{Eq:dual problem for convex modification}) satisfies $\mathrm{P}_q^* (\hat{\boldsymbol{\nu}}') \le \lambda$.
\end{proof}

\renewcommand{\theequation}{B-\arabic{equation}}
\setcounter{equation}{0}  
\section*{Appendix B: Proof of Lemma~\ref{lemma:max lambda}}

\noindent  \textbf{Proof of Lemma~\ref{lemma:max lambda}:}
\begin{proof}
  Since $\mathbf{D}$ is not of full rank by Lemma~\ref{lemma:properties of D}\ref{(i)}, the solution to (\ref{Eq:convex dual}) in the absence of constraint is not unique, and takes the form
\begin{equation}
\label{Eq:sol dual}
\begin{split}
\hat{\boldsymbol{\nu}} &=  (\mathbf{DD}^T)^{\dagger} \mathbf{Dx} + (\mathbf{I}-\mathbf{D}(\mathbf{D}^T\mathbf{D})^{\dagger}\mathbf{D}^T) \boldsymbol{\omega}\\
&= (\mathbf{D}^T)^{\dagger}\mathbf{x} + (\mathbf{I}-\mathbf{DD}^{\dagger})\boldsymbol{\omega}\\
&= \frac{1}{n} \mathbf{Dx} + (\mathbf{I}-\frac{1}{n}\mathbf{D}\mathbf{D}^T)\boldsymbol{\omega},
\end{split}
\end{equation}
for $\boldsymbol{\omega}\in \mathbb{R}^{\left[ p \cdot {n\choose 2}\right]}$.  The second equality follows from Lemma~\ref{lemma:properties of D}\ref{(iii)} and the last equality follows from Lemma~\ref{lemma:properties of D}\ref{(ii)}.

Let $\hat{\mathbf{u}}$ be the solution to (\ref{Eq:convex clustering2}).  Substituting $\hat{\boldsymbol{\nu}}$ given in (\ref{Eq:sol dual}) into (\ref{Eq:convex clustering2 solution}), we obtain
\begin{equation*}
\begin{split}
 \mathbf{D}\hat{\mathbf{u}} &= \mathbf{Dx}-\mathbf{DD}^T\hat{\boldsymbol{\nu}}\\
 &=  \mathbf{Dx}-\frac{1}{n}\mathbf{DD}^T \mathbf{Dx} - \mathbf{DD}^T\boldsymbol{\omega} + \frac{1}{n}\mathbf{DD}^T\mathbf{DD}^T \boldsymbol{\omega}\\
 &= \mathbf{Dx} - \mathbf{Dx} - \mathbf{DD}^T\boldsymbol{\omega} + \mathbf{DD}^T\boldsymbol{\omega}\\
 &= \mathbf{0}.
\end{split}
\end{equation*}
Recall from Definition~\ref{definition:cluster} that all observations are estimated to belong to the same cluster if $\mathbf{D}\hat{\mathbf{u}}=\mathbf{0}$.   For any $\hat{\boldsymbol{\nu}}$ in (\ref{Eq:sol dual}), picking $\lambda = \mathrm{P}^*_q (\hat{\boldsymbol{\nu}})$ guarantees that the constraint on the dual problem (\ref{Eq:convex dual}) is inactive, and therefore that convex clustering has a trivial solution of $\mathbf{D}\hat{\mathbf{u}}=\mathbf{0}$.  

Since $\hat{\boldsymbol{\nu}}$ is not unique, $\mathrm{P}^*_q (\hat{\boldsymbol{\nu}})$ is not unique.  
In order to obtain the smallest tuning parameter $\lambda$ such that $\mathbf{D}\hat{\mathbf{u}}=\mathbf{0}$, we take
\[
\lambda_{\mathrm{upper}} := \underset{\boldsymbol{\omega}\in \mathbb{R}^{\left[p\cdot {n\choose 2}\right]}}{\min}   \mathrm{P}^*_q \left( \frac{1}{n}\mathbf{Dx} + \left(\mathbf{I} - \frac{1}{n}\mathbf{D}\mathbf{D}^T\right) {\boldsymbol{\omega}}\right).
\] 
Any tuning parameter $\lambda \ge \lambda_{\mathrm{upper}}$ results in an estimate for which all observations belong to a single cluster.  The proof is completed by recalling the definition of the dual norm $\mathrm{P}_q^*(\cdot)$ in (\ref{dual norm}).
\end{proof}

\renewcommand{\theequation}{C-\arabic{equation}}
\setcounter{equation}{0}  
\section*{Appendix C: Proof of Lemmas~\ref{lemma:prediction consistency}---\ref{lemma:prediction consistency l2}}
To prove Lemmas~\ref{lemma:prediction consistency} and \ref{lemma:prediction consistency l2}, we need a lemma on the tail bound for quadratic forms of independent sub-Gaussian random variables.   

\begin{lemma}
\label{lemma:hanson wright}
\citep{hanson1971bound} Let $\mathbf{z}$ be a vector of independent sub-Gaussian random variables with mean zero and variance $\sigma^2$. Let $\mathbf{M}$ be a symmetric matrix.  Then, there exists some constants $c_1,c_2>0$ such that for any $t>0$,
\[
\mathrm{Pr} \left(  \mathbf{z}^T \mathbf{Mz}   \ge t+ \sigma^2 \mathrm{tr}(\mathbf{M}) \right)  \le  \exp \left\{   - \min \left( \frac{ c_1t^2}{\sigma^4\|\mathbf{M}\|_{\mathrm{F}}} , \frac{c_2 t}{\sigma^2\|\mathbf{M}\|_\mathrm{sp}} \right)   \right\},
\]
where $\|\cdot \|_{\mathrm{F}}$ and $\|\cdot \|_{\mathrm{sp}}$ are the Frobenius norm and spectral norm, respectively.
\end{lemma}

In order to simplify our analysis, we start by reformulating (\ref{Eq:convex clustering2}) as in \citet{liu2013guaranteed}. Let  $\mathbf{D}=\mathbf{A}\Lambda\mathbf{V}_{\beta}^T$ be the \emph{singular value decomposition} of $\mathbf{D}$, where $\mathbf{A}\in \mathbb{R}^{\left[ p\cdot {n \choose 2}\right] \times p(n-1)}$, $\Lambda \in \mathbb{R}^{p(n-1)\times p(n-1)}$, and $\mathbf{V}_{\beta} \in \mathbb{R}^{np \times p(n-1)}$.  Construct $\mathbf{V}_{\alpha} \in \mathbb{R}^{np \times p}$ such that $\mathbf{V} = [\mathbf{V}_{\alpha},\mathbf{V}_{\beta} ] \in \mathbb{R}^{np \times np}$
is an orthogonal matrix, that is, $\mathbf{V}^T\mathbf{V} = \mathbf{VV}^T = \mathbf{I}$.  Note that $\mathbf{V}_{\alpha}^T  \mathbf{V}_{\beta}= 0$.

 Let $\boldsymbol{\beta} = \mathbf{V}_{\beta}^T\mathbf{u} \in \mathbb{R}^{p(n-1)}$ and $\boldsymbol{\alpha} = \mathbf{V}_{\alpha}^T \mathbf{u}\in \mathbb{R}^{p}$.  Also, let $\lambda' = \frac{\lambda}{np}$.  
Optimization problem (\ref{Eq:convex clustering2}) then becomes
\begin{equation}
\label{Eq:convex clustering3}
\underset{\boldsymbol{\alpha} \in \mathbb{R}^p, \boldsymbol{\beta}\in \mathbb{R}^{p(n-1)}}{\mathrm{minimize}}\; \frac{1}{2np} \|\mathbf{x} - \mathbf{V}_{\alpha} \boldsymbol{\alpha} -  \mathbf{V}_{\beta}\boldsymbol{\beta} \|^2 + \lambda' \mathrm{P}_q(\mathbf{Z} \boldsymbol{\beta}),
\end{equation}
where $\mathbf{Z} = \mathbf{A}\Lambda \in \mathbb{R}^{\left[p\cdot {n \choose 2}\right] \times p(n-1)}$.  Note that $\mathrm{rank}(\mathbf{Z})=p(n-1)$ and therefore, there exists a pseudo-inverse $\mathbf{Z}^{\dagger}\in \mathbb{R}^{p(n-1) \times \left[p\cdot {n \choose 2}\right]}$ such that $\mathbf{Z}^{\dagger}  \mathbf{Z} = \mathbf{I}$.   Recall from Section~\ref{Sec:intro} that the set $\mathcal{C}(i,i')$ contains the row indices of $\mathbf{D}$ such that $\mathbf{D}_{\mathcal{C}(i,i')} \mathbf{u} = \mathbf{U}_{i.}-\mathbf{U}_{i'.}$.  Let the submatrices $\mathbf{Z}_{\mathcal{C}{(i,i')}}$ and $\mathbf{Z}^{\dagger}_{\mathcal{C}(i,i')}$ denote the rows of $\mathbf{Z}$ and the columns of $\mathbf{Z}^{\dagger}$, respectively, corresponding to the indices in the set $\mathcal{C}(i,i')$.  By Lemma~\ref{lemma:properties of D}\ref{(v)}, 
\begin{equation}
\small
\label{Eq:eigenvalue}
\Lambda_{\min} (\mathbf{Z}) = \Lambda_{\min} (\mathbf{D})= \frac{1}{\Lambda_{\max} (\mathbf{Z}^{\dagger})} = \sqrt{n} \qquad \text{and} \qquad \Lambda_{\max} (\mathbf{Z}) = \Lambda_{\max} (\mathbf{D})= \frac{1}{\Lambda_{\min} (\mathbf{Z}^{\dagger})} = \sqrt{n}.
\end{equation}
Let  $\hat{\boldsymbol{\alpha}}$ and $\hat{\boldsymbol{\beta}}$ denote the solution to (\ref{Eq:convex clustering3}).\\

\noindent \textbf{Proof of Lemma~\ref{lemma:prediction consistency}:}
\begin{proof}
We establish a finite sample bound for the prediction error of convex clustering with $q=1$ by analyzing (\ref{Eq:convex clustering3}).  First, note that $\hat{\mathbf{u}} = \mathbf{V}_{\alpha}\hat{\boldsymbol{\alpha}}+\mathbf{V}_{\beta}\hat{\boldsymbol{\beta}}$ and ${\mathbf{u}} = \mathbf{V}_{\alpha}{\boldsymbol{\alpha}}+\mathbf{V}_{\beta}{\boldsymbol{\beta}}$. Thus,  $\frac{1}{2np}\|\hat{\mathbf{u}}-\mathbf{u}\|^2 = \frac{1}{2np}\|\mathbf{V}_{\alpha}(\hat{\boldsymbol{\alpha}}-\boldsymbol{\alpha})+ \mathbf{V}_{\beta}(\hat{\boldsymbol{\beta}}-\boldsymbol{\beta})   \|^2$.  
Recall from (\ref{q norm}) that $\mathrm{P}_1(\mathbf{Z}\boldsymbol{\beta}) = \|\mathbf{Z}\boldsymbol{\beta}\|_1$.  By the definition of $\hat{\boldsymbol{\alpha}}$ and $\hat{\boldsymbol{\beta}}$, we have  
\begin{equation*}
\frac{1}{2np}\|\mathbf{x}-(\mathbf{V}_{\alpha} \hat{\boldsymbol{\alpha}}+ \mathbf{V}_{\beta} \hat{\boldsymbol{\beta}})\|^2 + \lambda' \|\mathbf{Z}\hat{\boldsymbol{\beta}}\|_1
\le \frac{1}{2np}\|\mathbf{x}-(\mathbf{V}_{\alpha} \boldsymbol{\alpha}+ \mathbf{V}_{\beta} \boldsymbol{\beta})\|^2 +\lambda'  \|\mathbf{Z}\boldsymbol{\beta}\|_1,
\end{equation*}
implying
\begin{equation}
\label{Eq:basic inequality}
\frac{1}{2np} \| \mathbf{V}_{\alpha}(\hat{\boldsymbol{\alpha}}-\boldsymbol{\alpha})+ \mathbf{V}_{\beta}(\hat{\boldsymbol{\beta}}-\boldsymbol{\beta})  \|^2 +\lambda' \|\mathbf{Z}\hat{\boldsymbol{\beta}}\|_1
\le  \frac{1}{np}\mathrm{G} (\hat{\boldsymbol{\alpha}},\hat{\boldsymbol{\beta}})+\lambda' \|\mathbf{Z}\boldsymbol{\beta}\|_1,
\end{equation}
where $\mathrm{G} (\hat{\boldsymbol{\alpha}},\hat{\boldsymbol{\beta}})=  \boldsymbol{\epsilon}^T \left[ \mathbf{V}_{\alpha}(\hat{\boldsymbol{\alpha}}-\boldsymbol{\alpha})+ \mathbf{V}_{\beta}(\hat{\boldsymbol{\beta}}-\boldsymbol{\beta}) \right] $.  Recall that $\mathbf{V}_{\alpha}^T\mathbf{V}_{\alpha}=\mathbf{I}$ and $\mathbf{V}_{\alpha}^T\mathbf{V}_{\beta} = \mathbf{0}$. By the optimality condition of (\ref{Eq:convex clustering3}), 
\begin{equation*}
\begin{split}
\hat{\boldsymbol{\alpha}} &= \mathbf{V}_{\alpha}^T  (\mathbf{x}-  \mathbf{V}_{\beta} \hat{\boldsymbol{\beta}} )\\
&=  \mathbf{V}_{\alpha}^T \left(  \mathbf{V}_{\alpha} \boldsymbol{\alpha} + \mathbf{V}_{\beta}\boldsymbol{\beta} + \boldsymbol{\epsilon} - \mathbf{V}_\beta \hat{\boldsymbol{\beta}}    \right)\\
&= \boldsymbol{\alpha} + \mathbf{V}^T_{\alpha} \boldsymbol{\epsilon}.
\end{split}
\end{equation*}
Therefore, substituting $\hat{\boldsymbol{\alpha}}-\boldsymbol{\alpha} = \mathbf{V}^T_{\alpha}\boldsymbol{\epsilon} $ into  $\mathrm{G}(\hat{\boldsymbol{\alpha}},\hat{\boldsymbol{\beta}})$, we obtain 
\begin{equation*}
\small
\begin{split}
\frac{1}{np}\left|\mathrm{G}(\hat{\boldsymbol{\alpha}},\hat{\boldsymbol{\beta}})\right| &=\frac{1}{np} \left|  \boldsymbol{\epsilon}^T \left[ \mathbf{V}_{\alpha}(\hat{\boldsymbol{\alpha}}-\boldsymbol{\alpha})+ \mathbf{V}_{\beta}(\hat{\boldsymbol{\beta}}-\boldsymbol{\beta}) \right] \right|\\
&=  \frac{1}{np}\left|   \boldsymbol{\epsilon}^T  \mathbf{V}_{\alpha} \mathbf{V}^T_{\alpha} \boldsymbol{\epsilon} + \boldsymbol{\epsilon}^T\mathbf{V}_{\beta}(\hat{\boldsymbol{\beta}}-\boldsymbol{\beta})\right|\\
&\le  \frac{1}{np}\boldsymbol{\epsilon}^T \mathbf{V}_{\alpha}\mathbf{V}_{\alpha}^T \boldsymbol{\epsilon} + \frac{1}{np}\left|\boldsymbol{\epsilon}^T  \mathbf{V}_\beta (\hat{\boldsymbol{\beta}}-\boldsymbol{\beta})  \right| \\
&= \frac{1}{np}\boldsymbol{\epsilon}^T \mathbf{V}_{\alpha}\mathbf{V}_{\alpha}^T \boldsymbol{\epsilon} +\frac{1}{np} \left|  \boldsymbol{\epsilon}^T \mathbf{V}_\beta \mathbf{Z}^\dagger \mathbf{Z} (\hat{\boldsymbol{\beta}}-\boldsymbol{\beta})   \right| \\
&\le  \frac{1}{np}\boldsymbol{\epsilon}^T \mathbf{V}_{\alpha}\mathbf{V}_{\alpha}^T \boldsymbol{\epsilon}+\frac{1}{np}\| \boldsymbol{\epsilon}^T \mathbf{V}_\beta \mathbf{Z}^\dagger\|_\infty \|\mathbf{Z} (\hat{\boldsymbol{\beta}}-\boldsymbol{\beta})  \|_1. \\
\end{split}
\end{equation*}
  We now establish bounds for $  \frac{1}{np}\boldsymbol{\epsilon}^T \mathbf{V}_{\alpha}\mathbf{V}_{\alpha}^T \boldsymbol{\epsilon}$ and $\frac{1}{np}\| \boldsymbol{\epsilon}^T \mathbf{V}_\beta \mathbf{Z}^\dagger\|_\infty$ that hold with high probability.\\
  
\noindent \textbf{Bound for} $\frac{1}{np}\boldsymbol{\epsilon}^T \mathbf{V}_{\alpha}\mathbf{V}_{\alpha}^T \boldsymbol{\epsilon}$:

First, note that $\mathbf{V}_{\alpha} \mathbf{V}_{\alpha}^T$ is a projection matrix of rank $p$, and therefore $\|\mathbf{V}_{\alpha}\mathbf{V}_{\alpha}^T\|_{\mathrm{sp}} = 1$ and $\|\mathbf{V}_{\alpha}\mathbf{V}_{\alpha}^T\|_{\mathrm{F}} = p$.  By Lemma~\ref{lemma:hanson wright} and taking $\mathbf{z} = \boldsymbol{\epsilon}$ and $\mathbf{M}=\mathbf{V}_{\alpha} \mathbf{V}_{\alpha}^T $, we have that 
\[
\mathrm{Pr}\left( \boldsymbol{\epsilon}^T \mathbf{V}_{\alpha}\mathbf{V}_{\alpha}^T \boldsymbol{\epsilon} \ge t + \sigma^2 p \right) \le  \exp \left\{  -\min \left( \frac{c_1 t^2}{\sigma^4 p}, \frac{c_2 t }{\sigma^2} \right) \right\},
\]
where $c_1$ and $c_2$ are constants in Lemma~\ref{lemma:hanson wright}.  Picking $t = \sigma^2 \sqrt{p\log (np)}$, we have 
\begin{equation}
\label{Eq:error bound chi square}
\mathrm{Pr}\left(\frac{1}{np} \boldsymbol{\epsilon}^T \mathbf{V}_{\alpha}\mathbf{V}_{\alpha}^T \boldsymbol{\epsilon} \ge  \sigma^2 \left[ \frac{1}{n}  + \sqrt{\frac{\log (np)}{n^2p}}  \right] \right) \le  \exp \left\{  -\min \left( c_1 \log (np), c_2 \sqrt{p\log (np)} \right) \right\}.
\end{equation}

\noindent \textbf{Bound for} $\frac{1}{np}\| \boldsymbol{\epsilon}^T \mathbf{V}_\beta \mathbf{Z}^\dagger\|_\infty$:

 Let $e_j$ be a vector of length $p\cdot {n \choose 2}$ with a one in the $j$th entry and zeroes in the remaining entries.  Let $v_j = e_j^T    (\mathbf{Z}^{\dagger})^T \mathbf{V}_{\beta}^T    \boldsymbol{\epsilon}$.  
  Using the fact that $\Lambda_{\max}(\mathbf{V}_{\beta}) = 1$ and $\Lambda_{\max}(\mathbf{Z}^{\dagger})= \frac{1}{\sqrt{n}}$ (\ref{Eq:eigenvalue}), we know that each $v_j$  is a  sub-Gaussian random variable with zero mean and variance at most 
$\frac{\sigma^2}{n}$. 
Therefore, by the union bound,
\[
\mathrm{Pr} \left( \underset{j}{\max} \; |v_j| \ge z    \right) \le p \cdot {n \choose 2} \cdot \mathrm{Pr}\left(|v_j|\ge z \right) \le 2 p \cdot {n \choose 2} \exp \left( -\frac{n z^2}{2\sigma^2} \right). 
\]

Picking $z= 2\sigma \sqrt{\frac{\log \left( p\cdot {n\choose 2}\right)}{n}}$, we obtain
\begin{equation}
\label{Eq:prediction consistency results bound}
\mathrm{Pr}\left( \|\boldsymbol{\epsilon}^T \mathbf{V}_{\beta} \mathbf{Z}^{\dagger}\|_{\infty} \ge 2\sigma \sqrt{\frac{\log (p\cdot {n \choose 2})}{n}}  \right)\le   \frac{2}{p \cdot {n \choose 2}}.
\end{equation}

\noindent \textbf{Combining the two upper bounds:}
Setting $\lambda' > 4\sigma \sqrt{\frac{\log \left( p\cdot {n \choose 2} \right)}{n^3p^2}}$
and combining the results from  (\ref{Eq:error bound chi square}) and (\ref{Eq:prediction consistency results bound}), we obtain
\begin{equation}
\label{Eq:bounds for prediction}
\frac{1}{np}\mathrm{G}(\hat{\boldsymbol{\alpha}},\hat{\boldsymbol{\beta}}) \le  \sigma^2  \left[ \frac{1}{n} + \sqrt{\frac{ \log (np)}{n^2p}} \right] + \frac{\lambda'}{2} \|\mathbf{Z}(\hat{\boldsymbol{\beta}}-\boldsymbol{\beta})\|_1 
\end{equation}
with probability at least $1-\frac{2}{p\cdot {n \choose 2}}-  \exp \left\{  -\min \left( c_1 \log (np), c_2 \sqrt{p\log (np)} \right) \right\}$.  Substituting (\ref{Eq:bounds for prediction}) into (\ref{Eq:basic inequality}), we obtain 
\begin{equation*}
\small
\begin{split}
&\quad \frac{1}{2np} \| \mathbf{V}_{\alpha}(\hat{\boldsymbol{\alpha}}-\boldsymbol{\alpha})+ \mathbf{V}_{\beta}(\hat{\boldsymbol{\beta}}-\boldsymbol{\beta})  \|^2 +\lambda' \|\mathbf{Z}\hat{\boldsymbol{\beta}}\|_1 \\
&\le \sigma^2  \left[ \frac{1}{n} + \sqrt{\frac{ \log (np)}{n^2p}}  \right] + \frac{\lambda'}{2} \|\mathbf{Z}(\hat{\boldsymbol{\beta}}-\boldsymbol{\beta})\|_1+  \lambda' \|\mathbf{Z}\boldsymbol{\beta}\|_1.
\end{split}
\end{equation*}
We get Lemma~\ref{lemma:prediction consistency} by an application of the triangle inequality and by rearranging the terms.
\end{proof}

\noindent \textbf{Proof of Lemma~\ref{lemma:prediction consistency l2}:}
\begin{proof}
We establish a finite sample bound for the prediction error of convex clustering with $q=2$ by analyzing  (\ref{Eq:convex clustering3}).  Recall from (\ref{q norm}) that $\mathrm{P}_2(\mathbf{Z}\boldsymbol{\beta}) = \sum_{i<i'} \|\mathbf{Z}_{\mathcal{C}(i,i')}\boldsymbol{\beta}\|_2$.  By the definition of $\hat{\boldsymbol{\alpha}}$ and $\hat{\boldsymbol{\beta}}$, we have  
\begin{equation*}
\frac{1}{2np}\|\mathbf{x}-(\mathbf{V}_{\alpha} \hat{\boldsymbol{\alpha}}+ \mathbf{V}_{\beta} \hat{\boldsymbol{\beta}})\|^2 + \lambda' \sum_{ i<i'}\|\mathbf{Z}_{\mathcal{C}(i,i')}\hat{\boldsymbol{\beta}}\|_2
\le \frac{1}{2np}\|\mathbf{x}-(\mathbf{V}_{\alpha} \boldsymbol{\alpha}+ \mathbf{V}_{\beta} \boldsymbol{\beta})\|^2 +  \lambda' \sum_{i<i'} \|\mathbf{Z}_{\mathcal{C}(i,i')}\boldsymbol{\beta}\|_2,
\end{equation*}
implying
\begin{equation}
\label{Eq:basic inequality l2}
\frac{1}{2np} \| \mathbf{V}_{\alpha}(\hat{\boldsymbol{\alpha}}-\boldsymbol{\alpha})+ \mathbf{V}_{\beta}(\hat{\boldsymbol{\beta}}-\boldsymbol{\beta})  \|^2 +\lambda'  \sum_{i<i'} \|\mathbf{Z}_{\mathcal{C}(i,i')}\hat{\boldsymbol{\beta}}\|_2
\le\frac{1}{np}  \mathrm{G} (\hat{\boldsymbol{\alpha}},\hat{\boldsymbol{\beta}})+\lambda' \sum_{ i<i'} \|\mathbf{Z}_{\mathcal{C}(i,i')}\boldsymbol{\beta}\|_2,
\end{equation}
where $\mathrm{G} (\hat{\boldsymbol{\alpha}},\hat{\boldsymbol{\beta}})=  \boldsymbol{\epsilon}^T \left[ \mathbf{V}_{\alpha}(\hat{ \boldsymbol{\alpha}}-\boldsymbol{\alpha})+ \mathbf{V}_{\beta}(\hat{\boldsymbol{\beta}}-\boldsymbol{\beta}) \right] $.  Again, by the optimality condition of (\ref{Eq:convex clustering3}), we have that $\hat{\boldsymbol{\alpha}}-\boldsymbol{\alpha} = \mathbf{V}^T_{\alpha}\boldsymbol{\epsilon} $.  Substituting this into  $\frac{1}{np}\mathrm{G}(\hat{\boldsymbol{\alpha}},\hat{\boldsymbol{\beta}})$, we obtain 
\begin{equation*}
\small
\begin{split}
\frac{1}{np}\left|\mathrm{G}(\hat{\boldsymbol{\alpha}},\hat{\boldsymbol{\beta}})\right| &= \frac{1}{np}\left|  \boldsymbol{\epsilon}^T \left[ \mathbf{V}_{\alpha}(\hat{\boldsymbol{\alpha}}-\boldsymbol{\alpha})+ \mathbf{V}_{\beta}(\hat{\boldsymbol{\beta}}-\boldsymbol{\beta}) \right] \right|\\
&=  \frac{1}{np}\left|   \boldsymbol{\epsilon}^T  \mathbf{V}_{\alpha} \mathbf{V}^T_{\alpha} \boldsymbol{\epsilon} + \boldsymbol{\epsilon}^T\mathbf{V}_{\beta}(\hat{\boldsymbol{\beta}}-\boldsymbol{\beta})\right|\\
&\le  \frac{1}{np}\boldsymbol{\epsilon}^T \mathbf{V}_{\alpha}\mathbf{V}_{\alpha}^T \boldsymbol{\epsilon} +\frac{1}{np} \left|\boldsymbol{\epsilon}^T  \mathbf{V}_\beta (\hat{\boldsymbol{\beta}}-\boldsymbol{\beta})  \right| \\
&=\frac{1}{np} \boldsymbol{\epsilon}^T \mathbf{V}_{\alpha}\mathbf{V}_{\alpha}^T \boldsymbol{\epsilon} +\frac{1}{np} \left|  \boldsymbol{\epsilon}^T \mathbf{V}_\beta \mathbf{Z}^\dagger \mathbf{Z} (\hat{\boldsymbol{\beta}}-\boldsymbol{\beta})   \right| \\
&=\frac{1}{np} \boldsymbol{\epsilon}^T \mathbf{V}_{\alpha}\mathbf{V}_{\alpha}^T \boldsymbol{\epsilon} +\frac{1}{np} \left|\sum_{i<i'}  (\boldsymbol{\epsilon}^T \mathbf{V}_\beta \mathbf{Z}_{\mathcal{C}(i,i')}^\dagger) (\mathbf{Z}_{\mathcal{C}(i,i')} (\hat{\boldsymbol{\beta}}-\boldsymbol{\beta}))   \right| \\
&\le\frac{1}{np} \boldsymbol{\epsilon}^T \mathbf{V}_{\alpha}\mathbf{V}_{\alpha}^T \boldsymbol{\epsilon} +\frac{1}{np} \sum_{i<i'}\left|  (\boldsymbol{\epsilon}^T \mathbf{V}_\beta \mathbf{Z}_{\mathcal{C}(i,i')}^\dagger) (\mathbf{Z}_{\mathcal{C}(i,i')} (\hat{\boldsymbol{\beta}}-\boldsymbol{\beta}))   \right| \\
&\le \frac{1}{np} \boldsymbol{\epsilon}^T \mathbf{V}_{\alpha}\mathbf{V}_{\alpha}^T \boldsymbol{\epsilon}+\frac{1}{np}\sum_{i<i'}\| \boldsymbol{\epsilon}^T \mathbf{V}_\beta \mathbf{Z}_{\mathcal{C}(i,i')}^\dagger\|_2 \|\mathbf{Z}_{\mathcal{C}(i,i')} (\hat{\boldsymbol{\beta}}-\boldsymbol{\beta})  \|_2 \\
&\le \frac{1}{np} \boldsymbol{\epsilon}^T \mathbf{V}_{\alpha}\mathbf{V}_{\alpha}^T \boldsymbol{\epsilon}+ \frac{1}{np}\cdot  \underset{i<i'}{\max} \;  \| \boldsymbol{\epsilon}^T \mathbf{V}_\beta \mathbf{Z}^\dagger_{\mathcal{C}(i,i')}\|_2 \sum_{ i<i'} \|\mathbf{Z}_{\mathcal{C}(i,i')} (\hat{\boldsymbol{\beta}}-\boldsymbol{\beta})  \|_2,\\
\end{split}
\end{equation*}
where the second inequality follows from an application of the triangle inequality and the third inequality from an application of  the Cauchy-Schwarz inequality.  
  We now establish bounds for $  \frac{1}{np}\boldsymbol{\epsilon}^T \mathbf{V}_{\alpha}\mathbf{V}_{\alpha}^T \boldsymbol{\epsilon}$ and $  \frac{1}{np} \cdot \underset{  i<i'}{\max} \;  \| \boldsymbol{\epsilon}^T \mathbf{V}_\beta \mathbf{Z}^\dagger_{\mathcal{C}(i,i')}\|_2$ that hold with large probability.\\

\noindent \textbf{Bound for} $\frac{1}{np}\boldsymbol{\epsilon}^T \mathbf{V}_{\alpha}\mathbf{V}_{\alpha}^T \boldsymbol{\epsilon}$:

This is established in the proof of Lemma~\ref{lemma:prediction consistency} in  (\ref{Eq:error bound chi square}), i.e., 
\[
\mathrm{Pr}\left( \frac{1}{np}\boldsymbol{\epsilon}^T \mathbf{V}_{\alpha}\mathbf{V}_{\alpha}^T \boldsymbol{\epsilon} \ge \sigma^2   \left[ \frac{1}{n} + \sqrt{\frac{ \log (np)}{n^2p}}   \right] \right)  \le  \frac{1}{np}.
\]

\noindent \textbf{Bound for} $\frac{1}{np}\cdot \underset{ i<i'}{\max} \;  \| \boldsymbol{\epsilon}^T \mathbf{V}_\beta \mathbf{Z}^\dagger_{\mathcal{C}(i,i')}\|_2$:

First, note that there are $p$ indices in each set $\mathcal{C}(i,i')$.  Therefore, 
\[
\| \boldsymbol{\epsilon}^T \mathbf{V}_\beta \mathbf{Z}^\dagger_{\mathcal{C}(i,i')}\|_2\le \sqrt{p} \cdot   \| \boldsymbol{\epsilon}^T \mathbf{V}_\beta \mathbf{Z}^\dagger_{\mathcal{C}(i,i')}\|_\infty.  
\]
Note that
\begin{equation}
\label{Eq:temp inequality}
\frac{1}{np}\cdot \underset{  i<i'}{\max} \;  \| \boldsymbol{\epsilon}^T \mathbf{V}_\beta \mathbf{Z}^\dagger_{\mathcal{C}(i,i')}\|_2 \le \sqrt{\frac{1}{n^2p}} \cdot  \underset{ i<i'}{\max} \; \| \boldsymbol{\epsilon}^T \mathbf{V}_\beta \mathbf{Z}^\dagger_{\mathcal{C}(i,i')}\|_\infty = \sqrt{\frac{1}{n^2p}} \cdot \| \boldsymbol{\epsilon}^T \mathbf{V}_\beta \mathbf{Z}^\dagger\|_\infty.
\end{equation}
Therefore, using (\ref{Eq:temp inequality}), 
\begin{equation}
\label{Eq:temp bound l2}
\begin{split}
&\quad\; \mathrm{Pr}\left(\frac{1}{np}\cdot \underset{ i<i'}{\max} \;  \| \boldsymbol{\epsilon}^T \mathbf{V}_\beta \mathbf{Z}^\dagger_{\mathcal{C}(i,i')}\|_2  \ge 2\sigma\sqrt{\frac{ \log \left(p \cdot {n \choose 2}\right)}{n^3p}}\right) \\
&\le\mathrm{Pr}\left(      \| \boldsymbol{\epsilon}^T \mathbf{V}_\beta \mathbf{Z}^\dagger\|_\infty        \ge 2\sigma\sqrt{\frac{\log \left( p\cdot {n \choose 2}\right)}{n}}  \right) \\
&\le \frac{2}{p \cdot {n\choose 2}},
\end{split}
\end{equation}
where the last inequality follows from (\ref{Eq:prediction consistency results bound}) in the proof of Lemma~\ref{lemma:prediction consistency}.

Therefore, for $\lambda' > 4\sigma \sqrt{\frac{ \log \left(p \cdot {n\choose 2}\right)}{n^3p}}$, we have $\frac{\lambda'}{2} <  \frac{1}{np}\cdot  \underset{  i<i'}{\max} \;  \| \boldsymbol{\epsilon}^T \mathbf{V}_\beta \mathbf{Z}^\dagger_{\mathcal{C}(i,i')}\|_2 $ with probability at most $\frac{2}{ p\cdot {n\choose 2}}$.  Combining the results from (\ref{Eq:error bound chi square}) and (\ref{Eq:temp bound l2}), we have that 
\begin{equation}
\label{Eq:bounds for prediction l2}
\frac{1}{np}\mathrm{G}(\hat{\boldsymbol{\alpha}},\hat{\boldsymbol{\beta}}) \le  \sigma^2  \left[ \frac{1}{n} + \sqrt{\frac{ \log (np)}{n^2p}}  \right] + \frac{\lambda'}{2} \sum_{ i<i'}\|\mathbf{Z}_{\mathcal{C}(i,i')}(\hat{\boldsymbol{\beta}}-\boldsymbol{\beta})\|_2 
\end{equation}
with probability at least $1-\frac{2}{p\cdot {n \choose 2}}-  \exp \left\{  -\min \left( c_1 \log (np), c_2 \sqrt{p\log (np)} \right) \right\}$.  Substituting (\ref{Eq:bounds for prediction l2}) into (\ref{Eq:basic inequality l2}) , we obtain 
\begin{equation*}
\small
\begin{split}
&\quad \; \frac{1}{2np} \| \mathbf{V}_{\alpha}(\hat{\boldsymbol{\alpha}}-\boldsymbol{\alpha})+ \mathbf{V}_{\beta}(\hat{\boldsymbol{\beta}}-\boldsymbol{\beta})  \|^2 +\lambda' \sum_{ i<i'}\|\mathbf{Z}_{\mathcal{C}(i,i')}\hat{\boldsymbol{\beta}}\|_2 \\
&\le \sigma^2  \left[ \frac{1}{n} + \sqrt{\frac{ \log (np)}{n^2p}}   \right] + \frac{\lambda'}{2} \sum_{ i<i'}\|\mathbf{Z}_{\mathcal{C}(i,i')}(\hat{\boldsymbol{\beta}}-\boldsymbol{\beta})\|_2+   \lambda'\sum_{ i<i'} \|\mathbf{Z}_{\mathcal{C}(i,i')}\boldsymbol{\beta}\|_2.
\end{split}
\end{equation*}
We get Lemma~\ref{lemma:prediction consistency l2} by an application of the triangle inequality and by rearranging the terms.
\end{proof}

\renewcommand{\theequation}{D-\arabic{equation}}
\setcounter{equation}{0}  
\section*{Appendix D: Proof of Lemma~\ref{lemma:dof:l2}}

\noindent \textbf{Proof of Lemma~\ref{lemma:dof:l2}:}
\begin{proof}  
Directly from the dual problem (\ref{Eq:convex dual}),  $\mathbf{D}^T\hat{\boldsymbol{\nu}}$ is the projection of $\mathbf{x}$ onto the convex set $K = \left\{  \mathbf{D}^T\boldsymbol{\nu} : \mathrm{P}^*_2 (\boldsymbol{\nu}) \le \lambda  \right\}$.  Using the primal-dual relationship $\hat{\mathbf{u}} = \mathbf{x}-\mathbf{D}^T \hat{\boldsymbol{\nu}}$, we see that $\hat{\mathbf{u}}$ is the residual from projecting $\mathbf{x}$ onto the convex set $K$.  By Lemma 1 of \citet{tibshirani2012degrees}, $\hat{\mathbf{u}}$ is continuous and almost differentiable with respect to $\mathbf{x}$.  
Therefore, by Stein's formula, the degrees of freedom can be characterized as $\mathrm{E}\left[ \mathrm{tr} \left( \frac{\partial \hat{\mathbf{u}}}{\partial \mathbf{x}}\right)  \right]$. 

Recall that $\mathbf{D}_{\mathcal{C}{(i,i')}}$ denotes the rows of $\mathbf{D}$ corresponding to the indices in the set $\mathcal{C}(i,i')$.  Let $\hat{\mathcal{B}}_2 = \{(i,i'): \|\mathbf{D}_{\mathcal{C}(i,i')} \hat{\mathbf{u}}\|_2 \ne 0\}$.
By the optimality condition of (\ref{Eq:convex clustering2}) with $q=2$, we obtain 
\begin{equation}
\label{Eq:dof temp 1}
(\mathbf{x}-\hat{\mathbf{u}}) = \lambda \sum_{i<i'} \mathbf{D}^T_{\mathcal{C}(i,i')} g_{\mathcal{C}(i,i')},
\end{equation}
where
\[
g_{\mathcal{C}(i,i')} =  \begin{cases}  \frac{\mathbf{D}_{\mathcal{C}(i,i')} \hat{\mathbf{u}} }{\|\mathbf{D}_{\mathcal{C}(i,i')} \hat{\mathbf{u}} \|_2} & \mathrm{if \;} (i,i')\in \hat{\mathcal{B}}_2. \\
\in \{\Gamma   : \|\Gamma\|_2 \le 1\} & \mathrm{if \;} (i,i')\notin \hat{\mathcal{B}}_2.
\end{cases}
\]
  We define the matrix $\mathbf{D}_{-\hat{\mathcal{B}}_2}$ by removing the rows of $\mathbf{D}$ that correspond to elements in $\hat{\mathcal{B}}_2$.   Let $\mathbf{P} = \left( \mathbf{I}- \mathbf{D}^T_{-\hat{\mathcal{B}}_2}  (\mathbf{D}_{-\hat{\mathcal{B}}_2}  \mathbf{D}^T_{-\hat{\mathcal{B}}_2})^{\dagger} \mathbf{D}_{-\hat{\mathcal{B}}_2}     \right)$ be the projection matrix onto the complement of the space spanned by the rows of $\mathbf{D}_{-\hat{\mathcal{B}}_2}$.

By the definition of $\mathbf{D}_{-\hat{\mathcal{B}}_2}$, we obtain $\mathbf{D}_{-\hat{\mathcal{B}}_2} \hat{\mathbf{u}} = \mathbf{0}$.  Therefore, $\mathbf{P}\hat{\mathbf{u}}=\hat{\mathbf{u}}$. Multiplying $\mathbf{P}$ onto both sides of (\ref{Eq:dof temp 1}), we obtain
\begin{equation}
\label{Eq:P dof l2}
\begin{split}
\mathbf{Px} - \hat{\mathbf{u}} &= \lambda \mathbf{P} \sum_{i<i'} \mathbf{D}^T_{\mathcal{C}(i,i')} g_{\mathcal{C}(i,i')}  \\
&=\lambda \mathbf{P} \sum_{(i,i')\in \hat{\mathcal{B}}_2} \frac{\mathbf{D}^T_{\mathcal{C}(i,i')} \mathbf{D}_{\mathcal{C}(i,i')}   \hat{\mathbf{u}}}{\|\mathbf{D}_{\mathcal{C}(i,i')}   \hat{\mathbf{u}}\|_2},
\end{split}
\end{equation}
where the second equality follows from the fact that $\mathbf{P} \mathbf{D}^T_{\mathcal{C}(i,i')} = \mathbf{0}$ for any $(i,i') \notin \hat{\mathcal{B}}_{2}$. 

\citet{vaiter2014degrees} showed that there exists a neighborhood around almost every $\mathbf{x}$ such that the set $\hat{\mathcal{B}}_2$ is locally constant with respect to $\mathbf{x}$.
Therefore,  the derivative of (\ref{Eq:P dof l2}) with respect to $\mathbf{x}$ is
\begin{equation}
\label{Eq:optimum dof l2}
\mathbf{P} - \frac{\partial \hat{\mathbf{u}}}{\partial \mathbf{x}}= \lambda \mathbf{P} \sum_{(i,i')\in \hat{\mathcal{B}}_2 } \left( \frac{\mathbf{D}^T_{\mathcal{C}(i,i')} \mathbf{D}_{\mathcal{C}(i,i')} }{\|\mathbf{D}_{\mathcal{C}(i,i')}   \hat{\mathbf{u}}\|_2} -  \frac{\mathbf{D}^T_{\mathcal{C}(i,i')} \mathbf{D}_{\mathcal{C}(i,i')}   \hat{\mathbf{u}}\hat{\mathbf{u}}^T \mathbf{D}^T_{\mathcal{C}(i,i')} \mathbf{D}_{\mathcal{C}(i,i')}  }{\|\mathbf{D}_{\mathcal{C}(i,i')}   \hat{\mathbf{u}}\|_2^3}\right) \frac{\partial \hat{\mathbf{u}}}{\partial \mathbf{x}},
\end{equation}
using the fact that for any matrix $\mathbf{A}$ with $\|\mathbf{Av}\|_2\ne 0$,  $\frac{\partial}{\partial \mathbf{v}} \frac{\mathbf{A}^T\mathbf{A} \mathbf{v}}{ \|\mathbf{A}\mathbf{v}\|_2} =\frac{\mathbf{A}^T\mathbf{A} }{ \|\mathbf{A}\mathbf{v}\|_2}   - \frac{\mathbf{A}^T\mathbf{A} \mathbf{vv}^T \mathbf{A}^T\mathbf{A}}{ \|\mathbf{A}\mathbf{v}\|_2^3} $.

Solving (\ref{Eq:optimum dof l2}) for $\frac{\partial \hat{\mathbf{u}}}{\partial \mathbf{x}}$, we have
 \begin{equation}
 \label{dof l2 temp}
\frac{\partial \hat{\mathbf{u}}}{\partial \mathbf{x}}=  \left[\mathbf{I}+ \lambda \mathbf{P}  \sum_{(i,i')\in \hat{\mathcal{B}}_2 } \left( \frac{\mathbf{D}^T_{\mathcal{C}(i,i')} \mathbf{D}_{\mathcal{C}(i,i')} }{\|\mathbf{D}_{\mathcal{C}(i,i')}   \hat{\mathbf{u}}\|_2} -  \frac{\mathbf{D}^T_{\mathcal{C}(i,i')} \mathbf{D}_{\mathcal{C}(i,i')}   \hat{\mathbf{u}}\hat{\mathbf{u}}^T \mathbf{D}^T_{\mathcal{C}(i,i')} \mathbf{D}_{\mathcal{C}(i,i')}  }{\|\mathbf{D}_{\mathcal{C}(i,i')}   \hat{\mathbf{u}}\|_2^3}\right)\right]^{-1} \mathbf{P}.
\end{equation}
Therefore, an unbiased estimator of the degrees of freedom is of the form
\begin{equation*}
\mathrm{tr} \left( \frac{\partial \hat{\mathbf{u}}}{\partial \mathbf{x}}\right) = \mathrm{tr} \left(  \left[\mathbf{I}+ \lambda  \mathbf{P} \sum_{(i,i')\in \hat{\mathcal{B}}_2 } \left( \frac{\mathbf{D}^T_{\mathcal{C}(i,i')} \mathbf{D}_{\mathcal{C}(i,i')} }{\|\mathbf{D}_{\mathcal{C}(i,i')}   \hat{\mathbf{u}}\|_2} -  \frac{\mathbf{D}^T_{\mathcal{C}(i,i')} \mathbf{D}_{\mathcal{C}(i,i')}   \hat{\mathbf{u}}\hat{\mathbf{u}}^T \mathbf{D}^T_{\mathcal{C}(i,i')} \mathbf{D}_{\mathcal{C}(i,i')}  }{\|\mathbf{D}_{\mathcal{C}(i,i')}   \hat{\mathbf{u}}\|_2^3}\right)\right]^{-1} \mathbf{P}\right).
\end{equation*}
\end{proof}

\end{document}